\documentclass[aps,prl,reprint,superscriptaddress]{revtex4-2}
\usepackage[english]{babel}
\usepackage{amsmath,amssymb,graphicx,color,comment}
\usepackage[bookmarks=true,colorlinks,citecolor=blue,urlcolor=blue, linkcolor=blue]{hyperref}

\usepackage[usenames,dvipsnames]{xcolor}
\usepackage{graphicx}
\usepackage{braket}
\usepackage[normalem]{ulem}
\usepackage{enumerate}
\usepackage{enumitem}
\usepackage{soul}

\usepackage{cancel}
\usepackage{multirow}

\graphicspath{{Figures/}}

\newcommand{\be}{\begin{equation}}
\newcommand{\ee}{\end{equation}}
\def\bes{\begin{subequations}}
\def\esu{\end{subequations}}
\newcommand{\Den}{{\bf e}}
\newcommand{\Dp}{{\bf p}}

\newcommand{\cc}{{\tilde{c}}}

\newcommand{\Cite}[1]{\mbox{\cite{#1}}}
\newcommand{\Eone}{{\mathcal{E}_{\text{1D}}}}
\newcommand{\Ethree}{{\mathcal{E}_{\text{3D}}}}
\newcommand{\dd}{{\rm d}}

\newcommand{\eff}{{\text{eff}}}
\newcommand{\dr}{{\text{dr}}}

\begin{document}
\newcommand{\titleinfo}{Observing Bethe strings in an attractive Bose gas far from equilibrium}

\title{\titleinfo}

\author{Milena Horvath}
\thanks{These authors contributed equally to this work.}
\affiliation{Institut f{\"u}r Experimentalphysik und Zentrum f{\"u}r Quantenphysik, Universit{\"a}t Innsbruck, Technikerstra{\ss}e 25, Innsbruck, 6020, Austria}

\author{Alvise Bastianello}
\thanks{These authors contributed equally to this work.}
\affiliation{Technical University of Munich, TUM School of Natural Sciences, Physics Department, 85748 Garching, Germany}
\affiliation{Munich Center for Quantum Science and Technology (MCQST), Schellingstr. 4, 80799 M{\"u}nchen, Germany}

\author{Sudipta Dhar}
\thanks{These authors contributed equally to this work.}
\affiliation{Institut f{\"u}r Experimentalphysik und Zentrum f{\"u}r Quantenphysik, Universit{\"a}t Innsbruck, Technikerstra{\ss}e 25, Innsbruck, 6020, Austria}

\author{Rebekka Koch}
\affiliation{Institute of Physics and Institute for Theoretical Physics, University of Amsterdam, PO Box 94485, 1090 GL Amsterdam, The Netherlands}

\author{Yanliang Guo}
\affiliation{Institut f{\"u}r Experimentalphysik und Zentrum f{\"u}r Quantenphysik, Universit{\"a}t Innsbruck, Technikerstra{\ss}e 25, Innsbruck, 6020, Austria}

\author{Jean-Sébastien Caux}
\affiliation{Institute of Physics and Institute for Theoretical Physics, University of Amsterdam, PO Box 94485, 1090 GL Amsterdam, The Netherlands}

\author{Manuele Landini}
\affiliation{Institut f{\"u}r Experimentalphysik und Zentrum f{\"u}r Quantenphysik, Universit{\"a}t Innsbruck, Technikerstra{\ss}e 25, Innsbruck, 6020, Austria}

\author{Hanns-Christoph  N{\"a}gerl}\email{christoph.naegerl@uibk.ac.at}
\affiliation{Institut f{\"u}r Experimentalphysik und Zentrum f{\"u}r Quantenphysik, Universit{\"a}t Innsbruck, Technikerstra{\ss}e 25, Innsbruck, 6020, Austria}

\begin{abstract}
Bethe strings are bound states of constituent particles in a variety of interacting many-body one-dimensional (1D) integrable quantum models relevant to magnetism, nanophysics, cold atoms and beyond. As emergent fundamental excitations, they are predicted to collectively reshape observable equilibrium and dynamical properties. Small individual Bethe strings have recently been observed in quantum magnets and superconducting qubits. However, creating states featuring intermixtures of many, including large, strings remains an outstanding experimental challenge. Here, using nearly integrable ultracold Bose gases, we realize such intermixtures of Bethe strings out of equilibrium, by dynamically tuning interactions from repulsive to attractive. We measure the average binding energy of the strings, revealing the presence of bound states of more than six particles. We find further evidence for them in the momentum distribution and in Tan’s contact, connected to the correlated density. Our data quantitatively agree with predictions from generalized hydrodynamics (GHD). Manipulating intermixtures of Bethe strings opens new avenues for understanding quantum coherence, nonlinear dynamics and thermalization in strongly-interacting 1D systems.
\end{abstract}

\maketitle

\begin{figure*}[t!]
\renewcommand{\figurename}{FIG.}
	\includegraphics[width=\textwidth]{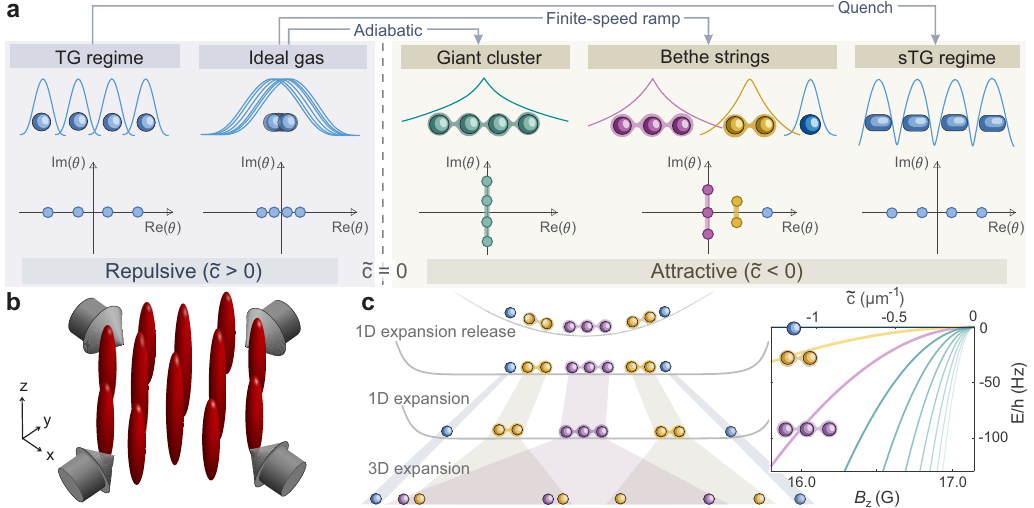}
    \caption{\textbf{Bethe strings and their detection in 1D Bose gases.} \textbf{a}, Pictorial representation of the different interaction regimes of 1D Bose gases. Wavefunction correlation of the different states (top panel) and corresponding rapidity distribution $\theta$ (bottom panel). As the interaction parameter $\cc$ is increased from $0$ (ideal gas) to $+\infty$ (TG regime) the gas fermionizes. The sTG regime can be accessed by quenching the interaction from the repulsive to the infinitely attractive regime. In contrast, adiabatically following the ground state of the system from positive to negative interactions $\cc$, and passing through $\cc\!=\!0$, results in a giant cluster. Sweeping $\cc$ from the strongly repulsive regime through the $\cc\!=\!0$ point with a finite rate realizes a non-equilibrium 1D gas composed of Bethe strings. Bound states with different number of particles are represented by different colours. \textbf{b}, Illustration of the experimental geometry. Horizontally propagating lattice beams (gray arrows) create an ensemble of independent 1D Bose gases (red tubes). \textbf{c}, Schematic of the 1D-3D expansion protocol used to probe Bethe strings.
    In the 1D the expansion, Bethe strings remain bound, breaking apart only upon the 3D release. Inset: Energy of different string states as a function of both $\cc$ and $B_z$, for bound states with up to ten particles.}
	\label{Fig_1}
\end{figure*}

Strong correlations present modern physics with both challenges and opportunities. 
In elementary cases, interactions merely renormalize bare excitations~\cite{schulz1995} or generate emergent collective hydrodynamic modes~\cite{Haldane1981JPhysC,Haldane1981PRL}.
A further (and far more exciting) possibility is for a non-perturbative overhaul of microscopic degrees of freedom to occur, invalidating any description of the many-body state from a non-interacting basis. The formation of bound states is a simple manifestation of this paradigm: Bardeen–Cooper–Schrieffer (BCS) pairs illustrate how bound states give rise to superconductivity~\cite{Cooper1956}, while quark confinement leads to the hadrons of high-energy physics~\cite{greensite2011}.
Bosonic systems are potentially far richer than fermionic ones, since they accommodate large compounds of tightly-bound and thus strongly-correlated emergent particles.
Realizing long-lived ensembles of attractive bosons, however, poses major practical challenges in analogy with the vacuum instability of attractive QED~\cite{Dyson1952}.
In bosonic systems, the absence of Pauli exclusion permits the formation of large complexes, which in turn collapse due to short-ranged molecular processes. Attractively interacting bosons have thus been observed only at weak interactions for a limited number of particles ~\cite{Gerton2000,Donley2001,Strecker2002}, or in one dimension for strong attractive interactions~\cite{Haller2009, Vladan2019,Kao2021,Yang2024Phantom,Chen2023} where bound states are suppressed due to energetics.
The realization of a system of strongly attractive bosons displaying hierarchies of large bound states has remained elusive due to these difficulties.

In 1D integrable quantum systems~\cite{takahashi2005}, attractive interactions do not necessarily induce collapse, but instead give rise to Bethe strings, multi-particle bound states predicted nearly a century ago~\cite{bethe1931theorie}. They emerge as stable excitations in spin chains~\cite{takahashi2005}, cold atomic gases~\cite{Guan2022}, and quantum magnets~\cite{Wang2018Experimental,Bera2020}. Their stability stems from the characteristic absence of diffractive collisions in integrable systems~\cite{takahashi2005}, which prevents bound states from decaying into lower-lying states. Recent studies show that large Bethe strings determine anomalous transport properties of integrable spin chains and drive the emergence of the Kardar-Parisi-Zhang (KPZ) universality class~\cite{Kardar1986,Ljubotina2017,Gopalakrishnan2023,Ilievski2021Superuniversality,Jepsen2020,Scheie2021,Wei2022,Rosenberg2024}. Beyond traditional condensed matter systems, they are also relevant in high-energy physics and string theory~\cite{Beisert2012}. A proper understanding of this wealth of manifestations requires the ability to create and manipulate dense intermixtures of Bethe strings of different sizes.

Bethe strings are expected to exist in the cold-atom Lieb-Liniger (LL) setting~\Cite{Lieb1963,McGuire1964}. Here, using ultracold attractively-interacting Cs atoms in 1D, we create a dense intermixture of strings by sweeping the interaction from repulsive to attractive, passing through the non-interacting point, as proposed in Ref.~\cite{Koch2021}. A fast and indiscriminate collapse of the gas is hindered by the approximate integrability~\Cite{Lieb1963,McGuire1964} of the system. Using a finite-rate sweep, we create a far-from-equilibrium state that exhibits multiple, including large, Bethe strings. Their typical size is set by the initial conditions, primarily the temperature and atomic cloud density. 
We detect their signature by comparing release measurements in 1D and 3D and by analyzing the momentum distribution and performing a measurement of Tan's contact~\cite{tan2008large}. 
Our data agrees well with the predictions~\cite{Bastianello2019,Koch2021} from generalized hydrodynamics~\cite{Alvaredo2016,Bertini2016}. 

Bethe strings arise as eigenstates of the celebrated LL Hamiltonian
\begin{equation}
\hat{H}_\text{LL}=-\frac{\hbar^2}{2m} \sum_i  \frac{\partial^2}{\partial z_i^2} +\frac{\hbar^2 \cc}{2m}\sum_{ i\ne j} \delta(z_i-z_j), 
\label{eq_ham}
\end{equation}
where the first term is the kinetic energy for the particles with mass $m$ and the second term models the inter-particle interactions with strength $\cc$. 
The eigenstates, obtained by Bethe ansatz~\cite{bethe1931theorie,Lieb1963}, are labeled by a set of rapidities $\theta$~\cite{takahashi2005}, which are generalized momenta of a set of emergent, stable quasiparticles.  Figure.~\ref{Fig_1}\textbf{a} shows a schematic representation of the different interaction regimes of the 1D Bose gas. For $\cc\!>\!0$, the particles interact repulsively and the rapidities are real quantities. For sufficiently strong repulsive interactions, the bosons fermionize in the Tonks-Girardeau (TG) state~\cite{Paredes2004,kinoshita2004}. A quench to strong attractive interactions ($\cc\!<\!0$) realizes an excited gaseous state, the super Tonks-Girardeau (sTG) state~\cite{Haller2009}. However, in the attractive regime, the rapidities are generally complex valued with non-zero imaginary parts. They are arranged in clusters that share the same real part $\lambda\equiv\text{Re}(\theta)$, symmetrically placed along the imaginary axis with spacing $|\cc|$. These states are the Bethe strings~\cite{Yajiang2006Ground,Calabrese2007}. An isolated Bethe string of size $n$ has wavefunction $\psi(z_1<...<z_n)\propto \exp\left(i\sum_{j=1}^n \theta_j z_j\right)$, with rapidities $\theta_j =\lambda+i |\cc|(n+1-2j)/2$. Its extent $\Delta z_n$ can be estimated from the wavefunction decay as $\Delta z_n=2[|\cc|(n-1)]^{-1}$, where the length scale is set by $|\cc|^{-1}$.
The total energy $E_n$ of the string is $E_\text{com;n}-E_\text{b;n}$, where $E_\text{com;n}=n\hbar^2~\lambda^2/(2m)$ is the center-of-mass energy and $E_\text{b;n}=\hbar^2\cc^2n(n^2-1)/(24 m)$ is the binding energy~\Cite{McGuire1964}.
The very ground state for $N$ particles is a giant Bethe state with $N$ constituents.
Bethe strings with a lesser number of constituents have to be realized in a non-equilibrium situation~\cite{Koch2021,Piroli2016Multiparticle,zill2018}, here realized with a finite-rate interaction ramp from the repulsive to the attractive phase, passing through the non-interacting point. The specifics of the initial state, the ramp rate, and the necessarily nonzero temperature will then determine the distribution of Bethe states in such a non-equilibrium situation.

Our experimental sequence starts with a 3D Bose-Einstein condensate (BEC) of Cs in a crossed-beam optical dipole trap, levitated against gravity by a magnetic force~\cite{Kraemer2004}. We control the atom number in the BEC within the range of $2\!\times\!10^4$ to $6\!\times\!10^4$, with an uncertainty of $10$\%, and the temperature between $10$ nK to $35$ nK, with a precision of $1$ nK. A broad Feshbach resonance allows us to tune the 3D s-wave scattering length $a_\text{3D}$ via an offset magnetic field $B_z$. Initially, $B_z$ is set to $20.8(1)$ G, at which $a_\text{3D}\!=\!209(4)$ $a_0$. The BEC is in the Thomas-Fermi regime. It is adiabatically loaded into two lattice beams that cross at $90^{\circ}$ in the horizontal $x-y$ plane (see Fig.~\ref{Fig_1}\textbf{b}). This results in approximately $4000$ isolated 1D tubes with trapping frequencies of $\omega_{\perp}/2\pi\!=\! 10.5(1)$ kHz in the transversal and $\omega_z/2\pi\!=\!29.3(1)$ Hz in the longitudinal direction for a lattice depth of $25$ $E_\text{r}$, where, $ E_\mathrm{r}\!= \!\pi^{2}\hbar^{2}/(2md^{2})$ is the recoil energy and $d\!=\!532.2$ nm is the lattice spacing set by the wavelength of the lattice light. This gives longitudinal and transversal harmonic oscillator lengths of $a_\text{ho}\!=\! 1.6(2)$ $\mu$m and $a_\perp\!=\!83(4)$ nm, respectively. We tune the average number of atoms per tube $N$ between $7$ and $30$ by varying the density of the initial BEC.
\begin{figure}[t!]
\renewcommand{\figurename}{FIG.}
\centering
	\includegraphics[width=\columnwidth]{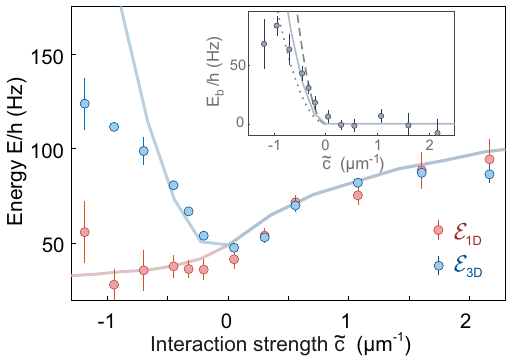}
	\caption{\textbf{Evidence for Bethe strings from release measurements.} Release energy $\Eone$ (red) and $\Ethree$ (blue) as a function of the interaction parameter $\cc$ at the end of the ramp. Each data point for the $\Eone$ ($\Ethree$) energy is determined from a fit of the change of the second moment of the cloud as a function of $t_\text{1D}$ ($t_\text{3D}$). The initial 1D temperature is estimated to be $T_\text{1D}\!=\!7$ nK~\cite{supp}, and $N = 11$. 
    Inset: The binding energy obtained via $E_\text{b} = \Ethree- \Eone$. 
    The dotted and dashed curves correspond to binding energy per particle of the Bethe strings of sizes $n=6$ and $10$ respectively. For both plots the standard error is given. The solid curves correspond to results from GHD.
    } 
    \label{Fig_2}
\end{figure}

In our 1D setting, the strength of the interaction is controlled via $\cc\!\simeq\!4a_\text{3D}/a^2_\perp$~\cite{Olshanii1998}.
After lattice loading, we set the interaction strength to $\cc\!=\!3.4(1)$ $\mu$m$^{-1}$, putting our systems in the moderately interacting TG regime. 
The 1D temperature $T_\text{1D}$ is set to values between $3$ nK and $35$ nK with an uncertainty of $20\%$~\cite{supp}.
We ramp the interaction parameter to lower and to negative values, crossing the non-interacting point, with a nearly constant rate of $\partial \cc/\partial t \approx -0.038$ $\mu$m$^{-1}$ms$^{-1}$. 
We expect that this procedure maps our system onto a distribution of Bethe strings as the interactions become attractive~\cite{Koch2021}.
For our experiment, for an interaction strength of $\cc\!=\!-1$ $\mu$m$^{-1}$, a string of size $n\!=\!4$ has a bond length of about $0.67$ $\mu$m. The binding energy of this string then is $h \times 190$ Hz $\approx k_\text{B} \times 9$ nK. 
Note that Bethe strings are only well defined when their bond length $\Delta z_n$ is much larger than $a_\perp$. For short distances, loss through short-range molecular processes is expected. Such loss is not captured by the LL model.

\begin{figure*}[t!]
\renewcommand{\figurename}{FIG.}
\centering
\includegraphics[width=\textwidth]{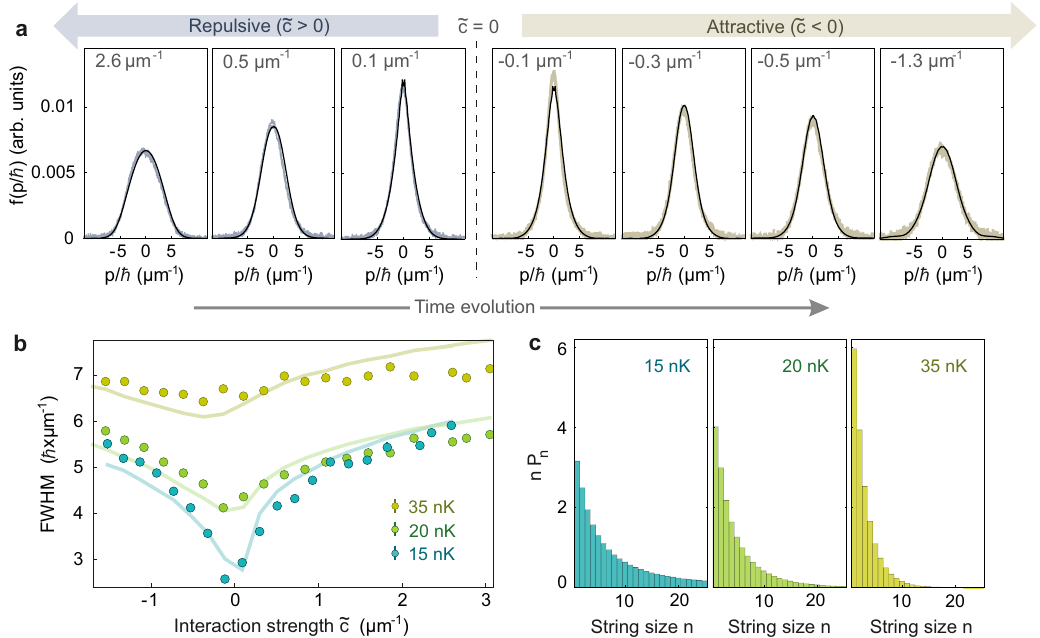}
	\caption{\textbf{Evidence for Bethe strings from the momentum distribution via a comparison with the results of GHD} \textbf{a}, Momentum distribution after 1D expansion $f(p/\hbar)$ for various values of the interaction parameter $\cc$ at the end of the ramp as indicated, from the repulsive (left) to the attractive (right) regime. Each distribution is the mean of five repetitions and the shaded region corresponds to the standard error. The data is compared with the results from GHD (solid line), which assumes an initial 1D temperature of $T_\text{1D}\!=\!15$ nK. \textbf{b}, FWHM of $f(p/\hbar)$ as a function of $\cc$ for three selected values of the $T_\text{1D}$ as indicated. Each data point is the average of five repetitions. The standard error is given. \textbf{c}, 
    Average number of atoms $nP_n$ in each string of size $n$ as predicted by theory for the temperatures as in Fig.~\ref{Fig_3}\textbf{b}.
	}
	\label{Fig_3}
\end{figure*}
We employ the rapidity measurement protocol~\cite{Wilson2020,Malvania2021,dubois2024,Li2023,Yang2024Phantom}, previously used in the absence of bound states, to probe our system. This protocol consists of a two-step expansion sequence: first, an expansion within the tubes for a time $t_\text{1D}$, followed by a free expansion into 3D space for time $t_\text{3D}$ (see Fig.~\ref{Fig_1}\textbf{c}). We ensure that $t_\text{1D}$ is large enough so that the system enters a dilute regime, where the interparticle interactions become negligible~\cite{supp}. In the repulsive regime, a subsequent absorption image yields the rapidity distribution of the gas. 
However, in the attractive regime the connection between the rapidity distribution and the absorption image becomes more intricate, as we explain below.
Experimentally, the expansion in the tubes is achieved by simultaneously switching off the dipole traps and removing the residual longitudinal confinement created by the lattice beams by applying a $808$-nm blue-detuned anti-trapping beam that flattens out the potential~\cite{Dhar2025}. For the second expansion step, the expansion into 3D, we simply shut off the transverse confinement and simultaneously set $a_\text{3D}\!\simeq\! 0$ to avoid any interaction-induced broadening of the density distribution of the expanding cloud. 

In the first experiment, we aim to find evidence of Bethe strings via measurements of the release energy. We extract the release energies $\Eone$ and $\Ethree$ from the evolution of the second moment of the expanding cloud during 1D and 3D phases of our rapidity measurement protocol. $\Eone$ is obtained by varying $t_\text{1D}$ while keeping $t_\text{3D}$ constant, and $\Ethree$ is obtained by varying $t_\text{3D}$ while keeping $t_\text{1D}$ constant~\cite{supp}. Figure~\ref{Fig_2} shows $\Eone$ and $\Ethree$ as a function of $\cc$. We find that for the repulsive case, $\Eone$ and $\Ethree$ match within the error, and decrease from about $h\times100$ Hz in the strongly repulsive side to approximately $h\times60$ Hz close to the non-interacting point. In contrast, a clear difference can be detected for attractive interactions. While $\Eone$ plateaus around $h\times35$ Hz, $\Ethree$ increases to about $h\times110$ Hz for $\cc\!=\!-1.0$ $\mu$m$^{-1}$. This discrepancy between $\Eone$ and $\Ethree$ for attractive interactions provides strong evidence for the presence of Bethe strings. During the 1D expansion, these bound states scatter, but remain intact, leading to a measured $\Eone$ that remains low. However, upon release into 3D, the sudden removal of the confinement breaks the bound states, converting their binding into kinetic energy, which leads to a larger release energy, mainly in the $z$-direction.
In the repulsive regime, on the other hand, the absence of bound states ensures that $\Eone$ and $\Ethree$ are equal. In fact, in the repulsive regime, $\Eone = \Ethree = U$, where $U=\langle \hat{H}_\text{LL}\rangle$ is the internal energy of the gas. In the attractive regime, $\Eone$ is given by the average center-of-mass energy of the strings $\Eone\!=\!\langle E_\text{com}\rangle$. In contrast, $\Ethree$ is the contribution from both the average center-of-mass energy of the strings and the binding energy $\Ethree\!=\!\langle E_\text{com}\rangle+\langle E_\text{b}\rangle$~\cite{supp}. The corresponding GHD simulations agree well with our measurements. In the inset of Fig.~\ref{Fig_2} we show the $E_\text{b}$ as a function of $\cc$ obtained via $\Ethree-\Eone$.
In order to account for the observed value of $\langle E_\text{b}\rangle$, the system necessarily contains strings of size $n^*$ or larger, where $n^*$ is the largest integer such that $\langle E_{\text{b}}\rangle \ge  E_{\text{b},n^*}/n^*$. We estimate $n^*$ to be six.

We find further evidence for Bethe strings by comparing the momentum distribution after 1D expansion $f(p/\hbar)$ with the predictions of GHD. For this we fix $t_\text{1D}\!=\!10$ ms and $t_\text{3D}\!=\!46.3$ ms. 
The momentum distribution is obtained from the longitudinal profile of the expanded cloud, converting position $z$ to momentum as $p=m z/(t_\text{1D}+t_\text{3D})$. 
In Fig.~\ref{Fig_3}\textbf{a} we show $f(p/\hbar)$ as the interaction strength $\cc$ is ramped from the repulsive to the attractive regime.  
At the beginning of the ramp $f(p/\hbar)$ is bell-shaped with a full width at half maximum (FWHM) $\hbar \times 6.0(1)$ $\mu$m$^{-1}$ 
As $\cc$ decreases and approaches the non-interacting point, $f(p/\hbar)$ becomes more sharply peaked, with the FWHM narrowing to $\hbar \times 3.0(1)$ $\mu$m$^{-1}$ at $\cc\!=\!0.1$ $\mu$m$^{-1}$. After crossing the zero-interaction point, $f(p/\hbar)$ broadens, nearly doubling in size by the end of the ramp at $\cc \! =\!-2.0$ $\mu\text{m}^{-1}$. 
Next we compare our data with GHD.
We estimate the best initial conditions for the simulations by optimizing the match with experimental data in the weakly repulsive regime, which is most sensitive to the initial parameters~\cite{supp}. We find $T_\text{1D} \!=\! 15$ nK.
The GHD simulations accounts for finite time-of-flight (TOF) effects by simulating the entire 1D and 3D expansions performed in the experiment~\cite{supp}. We find good agreement between the experimental data and the results from GHD. In the repulsive regime, $f(p/\hbar)$ corresponds to the rapidity distribution of the gas~\cite{supp}. Therefore, the narrowing of the distributions for decreasing $\cc>0$ is attributed to a reduction in the effective velocity of the quasiparticles. In contrast, for the attractive case $\cc<0$, the quasiparticles bind together to form Bethe strings that broaden the distributions. An intuitive understanding based on Heisenberg's uncertainty principle suggests that a string with bond length $\Delta z_n$ will release momentum within a window $\Delta p_{n} \sim \hbar/(2\Delta z_{n})$. This broadening effect becomes more prominent for larger Bethe strings and for stronger attractive interactions.

We now examine how the temperature affects the formation of the bound states. In Fig.~\ref{Fig_3}\textbf{b}, we show the FWHM of $f(p/\hbar)$ as a function of $\cc$ for three different values of $T_\text{1D}$. Note that for the $15$ nK dataset, we use a lattice depth of $20$ E$_\mathrm{r}$ instead of $25$ E$_\mathrm{r}$. As is already evident from Fig.~\ref{Fig_3}\textbf{a} the distribution broadens away from $\cc = 0$. This effect diminishes for higher $T_\text{1D}$. Results from GHD are in good agreement with our experimental data.
Figure~\ref{Fig_3}\textbf{c} shows the GHD prediction for the number of atoms $nP_n$ participating in the $n^{\text{th}}$ bound state. Here $P_n$ is the average number of strings of size $n$.
The probability of forming larger strings decreases with increasing $T_\text{1D}$. 
For increasing temperature, the kinetic energy of the system dominates over the interaction energy. This reduces the probability of forming larger bound states, and in turn reduces the average binding energy per particle. This is reflected in Fig.~\ref{Fig_3}\textbf{b}, where the FWHM of $f(p/\hbar)$ becomes less sensitive to the change of interaction at higher $T_\text{1D}$.  

The presence of Bethe strings significantly modifies the short-range correlations of the gas, notably the local pair correlations integrated over the trap $G_{2}$~\cite{Gangardt2003Stability,supp}, which quantify the probability of finding two particles in the same place.
This is directly proportional to Tan's contact $C=\cc^2 G_2$~\cite{tan2008large, tan2008energetics}.
Tan's contact establishes a fundamental link between microscopic quantities and the thermodynamic properties of the system. 
At thermal equilibrium, the virial theorem~\cite{tan2008generalized} relates $C$ to the internal energy $U$ and the potential energy $E_\text{V}$ of the harmonic trap as
\be\label{eq_virial}
C = \frac{2m \cc}{\hbar^2}(U - E_\text{V})\, .
\ee
We find that Eq.~\eqref{eq_virial} also generalizes to out-of-equilibrium scenarios described by stationary solutions for GHD~\cite{supp}.

\begin{figure}[t!]
\renewcommand{\figurename}{FIG.}
\centering
	\includegraphics[width=\columnwidth]{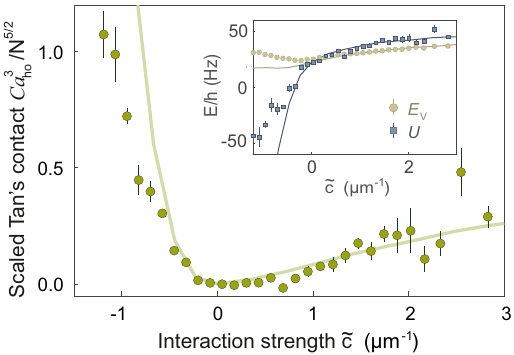}
	\caption{\textbf{Interaction dependence of Tan's contact.} 
    The scaled contact (circles) is extracted from experimental measurements of $U$ and $E_V$ via Eq.~\eqref{eq_virial}. Inset: The potential energy $E_\text{V}$ (circles) and internal energy $U$ (squares) as a function of $\cc$. The solid curves in both plots are results obtained from GHD. For these measurements $T_\text{1D} = 3$ nK and $N = 5$. The standard error is given.}
	\label{Fig_4}
\end{figure}
Experimentally, we obtain $U$ from the 3D release-energy measurements described previously. While for repulsive interactions we can directly extract $U$ from $\Ethree$, for attractive regime this is not the case. As shown in Fig.~\ref{Fig_2}, $E_\text{com}$ plateaus within experimental uncertainty for attractive interactions. We therefore approximate $E_\text{com}$ ($\cc<0$) $\simeq \Ethree$ ($\cc = 0$). The binding energy $\langle E_\text{b}\rangle$ is given by $\Ethree - \Ethree$ ($\cc = 0$). We use this approximation to extract $U$ for the attractive regime.
The potential energy is measured by imaging the spatial distribution of the atom cloud. We use a short TOF  
of $1.5$ ms without 1D expansion in order to lower the optical density before imaging. The potential energy per particle is given by $E_\text{V} = (1/2) m \omega_z^2 \langle z^2 \rangle$, where $\langle z^2 \rangle$ is the mean squared width of the cloud in the longitudinal direction. In Fig.~\ref{Fig_4} we plot the interaction dependence of the contact, determined from the measured $U$ and $E_\text{V}$ using Eq.~\eqref{eq_virial} (see inset of Fig.~\ref{Fig_4}).
On the repulsive side, Tan's contact decreases as $\cc$ is reduced. Upon entering the attractive regime, Tan's contact increases dramatically. It is nearly six times higher on the attractive side at the same magnitude of the interaction $|\cc|=1$ $\mu$m$^{-1}$. 
The GHD predictions qualitatively describe experimental observations with very good agreement on the repulsive side. For $\cc<0$, we attribute the discrepancy to finite-size effects, as theory predicts a substantial contribution from Bethe strings larger than the average number of particles in the tubes. Additional discrepancies stem from atom loss in the experiment, with approximately 20\% of the atoms being lost during the ramp for the strongest attractive interaction~\cite{supp}. Our data can be used to benchmark other theoretical methods for calculating Tan's contact in 1D~\cite{Vignolo2013,Lang2016Tans,Yao2018Tans}.

In summary, we have used 1D attractive Bose gases to realize novel non-equilibrium states of quantum matter featuring dense intermixtures of Bethe strings. In our experiment, approximate integrability stabilizes these bound states, and bypasses the rapid collapse normally associated to attractive bosons~\cite{Gerton2000,Donley2001,Strecker2002}. We have experimentally characterized these bound states through measurements of the binding energy, Tan’s contact, and the momentum distribution following 1D expansion. Our results agree with predictions from GHD, showcasing the latter's validity beyond the previously-considered repulsive~\cite{Schemmer2019,Malvania2021,Moller2021,Cataldini2022,dubois2024,Schuttelkopf2024} and sTG regimes~\cite{Yang2024Phantom}. Our results show that the presence of Bethe strings (somewhat similar to Efimov physics~\cite{Braaten2007}) in correlated gases greatly enriches the landscape of achievable non-equilibrium states in the vicinity of integrability.
Since Bethe strings display remarkable resilience to integrability breaking~\cite{morvan2022b}, they can alter transport properties~\cite{Bulchandani2021}, give rise to new interesting pre-thermal phases~\cite{Birnkammer2022} and affect the emergent hydrodynamics and thermalization of nearly-integrable systems~\cite{Caux2019,Cao2018,Bastianello2020Newton,Cataldini2022}. Going further, it would be interesting to study their participation in the emergence of KPZ universality~\cite{de2020delta} or their fate under dimensional crossover~\cite{Guo2024}.
\\

\noindent{\bf Data availability.} Experimental and simulation data are available on Zenodo~\cite{Zenodo}.
\\

\bigskip
\noindent{\bf Acknowledgments}\\
The Innsbruck team acknowledges funding by a Wittgenstein prize grant under the Austrian Science Fund's (FWF) project number Z336-N36, by the European Research Council (ERC) under project number 789017, by an FFG infrastructure grant with project number FO999896041, and by the FWF's COE 1 and quantA. Y.G. is supported by the FWF with project number 10.55776/COE1. MH thanks the doctoral school ALM for hospitality, with funding from the FWF under the project number W1259-N27. 
AB acknowledges support from the Deutsche Forschungsgemeinschaft (DFG, German Research Foundation) under Germany’s Excellence Strategy–EXC–2111–390814868. JSC and RK acknowledge support from the European Research Council under ERC Advanced grant 743032 DYNAMINT.
\\

\bibliography{biblio}

\begin{thebibliography}{78}
\bibitem [{Zen()}]{Zenodo}%
  \BibitemOpen
  \href@noop {} {}\bibinfo {howpublished} {{Research data are freely available from Zenodo at doi.org/10.5281/zenodo.15426226}}\BibitemShut {NoStop}%
\end{thebibliography}

\clearpage
\onecolumngrid
\newpage

\setcounter{equation}{0}  
\setcounter{figure}{0}  
\setcounter{page}{1}
\setcounter{section}{0} 
\renewcommand\thesection{\arabic{section}}  
\renewcommand\thesubsection{\arabic{subsection}}   
\renewcommand{\thetable}{S\arabic{table}}
\renewcommand{\theequation}{S\arabic{equation}}
\renewcommand{\thefigure}{S\arabic{figure}}
\setcounter{secnumdepth}{2}  

\begin{center}
{\large \textbf{Supplementary Materials of\\ ``\titleinfo''}}\\
\vspace{5pt}

Milena Horvath, Alvise Bastianello, Sudipta Dhar, Rebekka Koch, Yanliang Guo, \ \\
Jean-Sébastien Caux,
Manuele Landini, Hanns-Christoph  N{\"a}gerl
\end{center}

\section{Supplementary note 1: Estimating the populations of atoms in the tubes and fitting the initial conditions}
\label{sec:Sup_mat_Exp_prep}

Since the formation of Bethe strings in our protocol is a non-linear function of the density, having a good quantitative estimation of the distribution of atoms across the tubes is of key importance. Before the creation of the one-dimensional tubes, we assume the gas is at thermal equilibrium and it remains so until the dimensional cross-over: when the transverse trapping is strong enough, the tubes cannot exchange particles any longer and are effectively decoupled. Just before the tubes decouple, we approximate the gas as a collection of one-dimensional systems at thermal equilibrium, characterized by a unique temperature and a local chemical potential according to the shallow three dimensional trap~\cite{Meinert2015}. We also approximate the one-dimensional interactions $\tilde{c}$ to be the same as the one at the end of the cross-over in one dimension.
Below, we summarize in a table the relevant parameter for the tubes' loading.

\begin{figure*}[b!]
\centering

	\includegraphics[width=0.9\columnwidth]{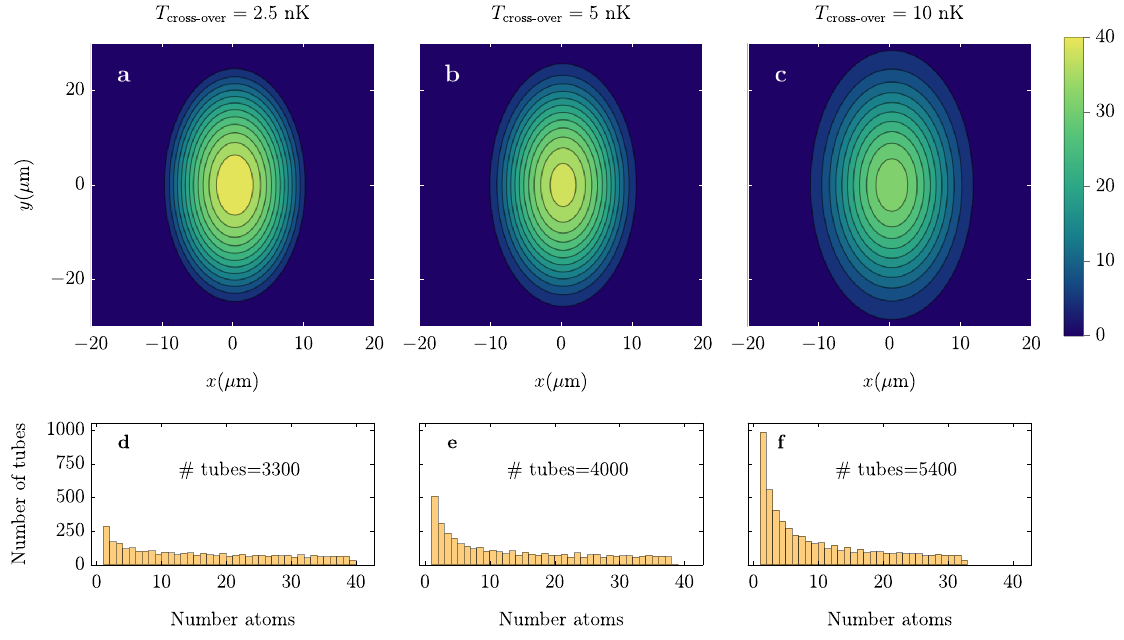}
	\caption{\textbf{Example of tubes' population varying $T_\text{cross-over}$.} As a concrete example, we show how tuning the dimensional crossover, the temperature changes the population of the tubes. In this example, we focus on the dataset of Fig.~\ref{Fig_3} corresponding to $T_\text{1D}=15\text{ nK}$. From left to right, the three columns corresponds to the choices $T_\text{cross-over}=(2.5,5,10)\text{ nK}$: the choice that best fits the initial data is $T_\text{cross-over}=5\text{ nK}$. In the first row (\textbf{a} to \textbf{c}), we show the spatial distribution of the tube population in the $(x,y)$ plane, whereas the longitudinal tubes are oriented in the $z$ direction. By increasing $T_\text{cross-over}$, more tubes are populated, but the average number of atoms in each of them decreases: this affects the creation of bound states when crossing to the attractive regime, as tubes with larger atom density populate larger bound states. In the second row (\textbf{d} to \textbf{f}), we show an histogram of the atom number population: tubes with less than one atom are not accounted for, and ``\# tubes" is the total the number of populated tubes in each case.
	}
	\label{Fig_SM_atms}
\end{figure*}

Under these assumptions, we can tabulate the distribution of atoms using thermodynamic Bethe ansatz (see Supplementary Note~\ref{sec_supp4}), keeping $T_\text{cross-over}$ as a tunable parameter. Notice that during the dimensional crossover, the trapping frequency along the $z-$ direction smoothly changes from the loading frequency $w_1$ to the final 1D value: the thermodynamics of the 1D tubes at the dimensional crossover is computed using the loading frequency as the frequency of the 1D trap.
An example of how different choices of $T_\text{cross-over}$ affect the tubes' population is provided in  Fig.~\ref{Fig_SM_atms}, a working code solving the thermodynamics of the one-dimensional tubes is provided on Zenodo~\cite{Zenodo}.
From the dimensional cross-over to the true 1D regime, the transverse trap is further increased, but the number of atoms in each tube remains constant. As the system enters deeper in the 1D regime, it approaches the integrable limit hindering thermalization. We assume that at the end of the lattice loading protocol, each tube is still well-approximated by a thermal ensemble with a new temperature $T_{1\text{D}}$, assumed to be uniform across all the tubes. The validity of this assumption is a posteriori supported by the good agreement with experimental data. 
The parameters $T_{1\text{D}}$ and $T_\text{cross-over}$ are estimated by matching as the experimental data in the weakly repulsive regime (see Supplementary Note~\ref{sec_supp4}). In Tab~\ref{tab:Exp_sim_parameters}, we summarize the experimental parameters and the corresponding results from the simulations for each of the datasets presented in the main text.

\begin{table}[t!]\label{tab:Exp_sim_parameters}
 \renewcommand{\tablename}{Tab.}
    \centering
  \begin{center}
\begin{tabular}{ p{1.0cm} c c c c c c c c }
 &\multicolumn{4}{c}{Experiment} & \multicolumn{4}{c}{Simulation}\\
 \hline
 &$\quad N_\text{BEC}\quad$ &$\quad(w_1,w_2,w_3)/2\pi\quad$& $\quad \omega_z/2\pi \quad$&  $\quad \omega_\perp/2\pi \quad$ &$\quad T_\text{cross-over}\quad$ & $\quad T_\text{1D}\quad$ &$N\quad$ &$N_\text{cent}\quad$\\ 
      &($\times 10^4$) &(Hz) &(Hz) &(kHz) &(nK) &(nK) &  &  \\ 
 \hline
Fig.~2& $3.2$& $(25.1,10.1,27.1)$& $29.3$& $10.5$& $10$& $7$& $11$& $21$\\
Fig.~3& $5.7$& $(25.1,10.1,27.0)$& $28.8$&$ 9.0$& $ 5$& $15$& $25$& $41$\\
      & $5.7$& $(25.1,10.1,27.0)$& $27.6$& $ 10.5$& $10$& $20$& $19$& $34$\\
      & $5.7$& $(25.1,10.1,27.0)$& $27.1$& $10.5$& $10$& $35$& $19$& $34$\\
Fig.~4& $1.9$& $(10.3,5.7,11.7)$& $18.1$& $11.5$& $3 $& $3 $& $5$& $9$ \\
\end{tabular}
 \caption{\textbf{Experimental and theoretical parameters for each dataset presented in the main text}. For the experimental parameters, we give the the initial atom number in the BEC $N_\text{BEC}$, the lattice loading frequencies $(\omega_1, \omega_2, \omega_3)/2\pi$ in the $z,x,y$ directions, respectively. After lattice loading, the longitudinal $\omega_z/2\pi$ and transversal $\omega_\perp/2\pi$ trapping frequencies in the tubes are also indicated. From the simulation, we estimate the cross-over temperature $T_\text{cross-over}$, the 1D temperature $T_\text{1D}$, and the average atom number per tube $N$ and the central tube atom number $N_\text{cent}$. }
 \end{center}
\end{table}

\section{Supplementary note 2: Summary of generalized hydrodynamics}
\label{sec:sup_mat_GHD}
Here we provide a compact summary of GHD, overviewing the main equations and how theory describes the momentum measurement after 1D expansion. A more detailed discussion for the interested reader can be found in Supplementary Note~\ref{sec_supp4}.
GHD~\cite{Alvaredo2016,Bertini2016} is a non-perturbative kinetic theory for integrable systems, governed at the Euler scale by the main equation 
\be
\label{eq_ghd}
\partial_t\rho_{n}(\lambda) + \partial_z[v^\eff_{n}(\lambda)\rho_{n}(\lambda)] +
\partial_\lambda[F^\eff_{n}(\lambda)\rho_{n}(\lambda)]=0\, .
\ee
The root density $\rho_{n}(\lambda)\to \rho_{n;t,z}(\lambda)$
is the phase-space density of Bethe strings of species $n$ in the position-rapidity plane. For $\cc>0$, $\rho_{n=1}$ is the only non-zero root density, since no bound states are present. The effective velocity $v_n^\eff$~\cite{Bertini2016,Alvaredo2016} is renormalized by interactions accounting for non-trivial scattering, while the weak integrability-breaking induced by the trap $\partial_zV\ne 0$~\cite{Doyon2017} and the effect of slow interaction changes $\cc\to \cc(t)$~\cite{Bastianello2019} are captured by the effective forces $F^\eff_n$. Ref.~\cite{Koch2021} connects the GHD equations across $\cc=0$ and determines the population of Bethe strings.
Within GHD, the internal and potential energies of the gas are $U=\sum_n\int \dd z\dd\lambda \, E_n(\lambda)\rho_n(\lambda)$ and $E_V=\sum_n\int \dd z\dd\lambda \, V(z) n\rho_n(\lambda)$ respectively.

The combined 1D and 3D expansion of our protocol can be also described within GHD. After reaching the dilute regime upon expanding in 1D with $F^\eff_n=0$, Bethe strings travel with their bare velocity $\hbar \lambda/m$.
Then, the transverse trap is removed and interactions quenched to zero. Particles released from a Bethe string acquire new momenta $p$ due to the converted binding energy, resuming the expansion and leading to a longitudinal density profile $d(z)$
\begin{equation}
\label{eq_SM_dz}
d(z)= \sum_n \int \dd \lambda \dd p\, \delta\left(z- \tfrac{ t_\text{1D}\hbar\lambda+t_\text{3D} p}{m}\right) C_n\left(\tfrac{p\hbar^{-1}-\lambda}{|\cc|}\right) \tfrac{n \bar{\rho}_{n}(\lambda)}{|\cc|} ,
\end{equation}
where $\bar{\rho}_n(\lambda)$ is the rapidity distribution integrated over the whole cloud.
$C_n$ are universal bell-shaped function describing the momentum distribution obtained from an isolated Bethe string quenched to zero interaction, see Supplementary Note~\ref{sec_supp4} for details.
For simplicity, in Eq.~\eqref{eq_SM_dz} we neglect the initial width of the cloud, and the short transient where the Bethe strings' velocity is renormalized by interactions: these effects are included simulating the 1D expansion with the GHD Eq.~\eqref{eq_ghd}.
In the repulsive phase, $d(z)$ maps to the rapidity distribution since $C_1(x)=\delta(x)$. This identification is lost in the attractive phase in the presence of Bethe strings. 
One can explicitly compute
$\int \dd z\, x^2 C_n(x)=n(n^2-1)/12$ (see Supplementary Note~\ref{eq_SM_dz}), leading to a simple expression for the variance of the expanded density profile
$\langle z^2\rangle\equiv \int \dd z\, z^2 d(z)\Big/\int \dd z\, d(z)$
\be \label{eq_second_moment}
\langle z^2\rangle=\frac{2(t_\text{1D}+t_\text{3D})^2}{m}\langle E_\text{com}\rangle+\frac{2 t_\text{3D}^2}{m}\langle E_\text{b}\rangle\, ,
\ee
which we use to extract the average center of mass and binding energies upon varying the 1D and 3D expansions.
The average of $E_\text{com}$ and $E_\text{b}$ is taken over the root densities, normalized to the particles number.

The pair correlation integrated over the trap is formally defined as $G_2\equiv \left\langle \sum_{i\ne j}\delta(z_i-z_j)\right\rangle$.
In Supplementary note~\ref{sec_supp4}, we give an explicit formula for $G_2$ in terms of the root densities~\eqref{eq_psi2}, and provide a more detailed overview of GHD and a discussion of finite expansion's time effects.

\section{Supplementary note 3: Determining the release energies of the system via the second moment}
\label{sec:sup_mat_Expansion_measurements}

\begin{figure}[t!]
\renewcommand{\figurename}{FIG.}
\centering
	\includegraphics[width=0.9\textwidth]{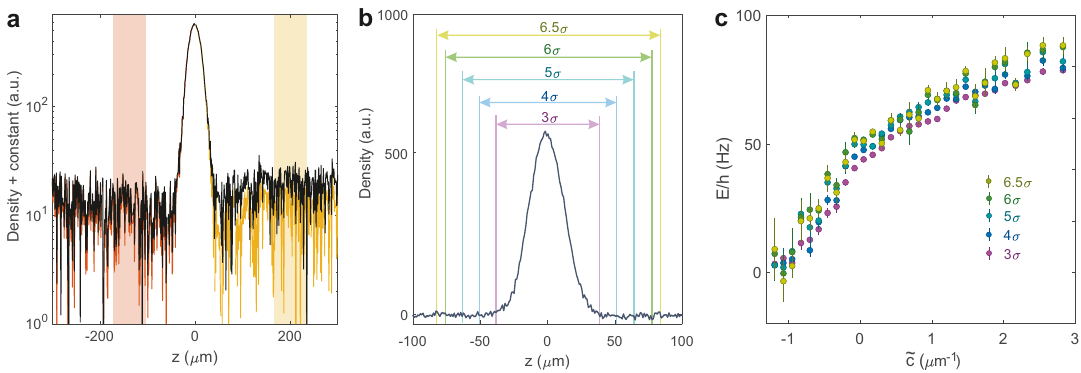}
	\caption{\textbf{Convergence of energy extracted from the second moment of the distribution.} \textbf{a}, Typical TOF profile of the cloud in log-linear scale. $z$ denotes the spatial position after TOF. For this plot, a small constant is added to TOF distribution in order to avoid negative values in logarithm. Different offsets are used on either side of the distribution in order to ensure that the second momentum converges for increasing regions of interest. Original data is shown in black and the colored curves correspond to the data after offset (mean over shaded region) has been applied. \textbf{b}, Typical TOF distribution indicating different region of interests used to obtain the internal energy $U$ shown in \textbf{c}. \textbf{c}, Internal energy determined using the different region of interest shown in \textbf{b}.}
    \label{Fig_SupMat_image_analysis}
\end{figure}

We determine the release energies $\Eone$ and $\Ethree$ by measuring the expansion rate of the second moment of the sample $\langle z^2\rangle$ during the 1D and 3D expansion stages, respectively. For the measurement shown in Fig.~\ref{Fig_2}, we scan $t_\text{1D}$ ($t_\text{3D}$) from $6$ to $11$ ms ($18$ to $43$ ms), while keeping $t_\text{3D} = 20.4$ ms ($t_\text{1D}\! =\! 8$ ms) fixed. For each experimental absorption image, we apply a background noise removal technique~\cite{Li2007} to improve the accuracy in the determination of the second moment. In Fig.~\ref{Fig_SupMat_image_analysis}\textbf{a} we show a typical TOF density profile after noise removal. We find that far away from the signal, the distribution has a non-zero offset. Furthermore, this offset is different on either side of the distribution, which we attribute to the inhomogeneous profile of the imaging beam. Although the difference between the two offset values is less than $1$\%, it is crucial for determining the second moment $\langle z^2\rangle$ of the distribution. In order to faithfully calculate the energy from $\langle z^2\rangle$, we use different regions of interest (see Fig.~\ref{Fig_SupMat_image_analysis}\textbf{b}) and check for convergence in the estimated energy. As shown in Fig.~\ref{Fig_SupMat_image_analysis}\textbf{c}, we see a convergence in energy above $3$$\sigma$ region of interest, where $\sigma$ is the gaussian width of the distribution. The energy is determined by fitting the evolution of $\langle z^2 \rangle$ as a function of expansion time. For the 1D release energy, we fit a parabolic function $a (t_\text{1D}+t_\text{fix})^2 +b$ to our 1D-expansion data, where $a$ and $b$ are fitting parameters, and $t_\text{fix}$ is the fixed 3D expansion time. The fitting parameter $a$ is related to $\Eone$ via $\Eone=2a/m$ as given in Eq.~\ref{eq_second_moment}. Similarly, for the 3D release energy, we use a parabolic fitting function $a (t_\text{3D}+t_\text{fix})^2 +b$ to our 3D-expansion data. Here, $t_\text{fix}$ is the fixed 1D expansion time. We extract $\Ethree$ from the fit via $\Ethree=2a/m$.

\section{Supplementary note 4: Atom loss and lifetime measurements}
\begin{figure*}[t!]
\centering
	\includegraphics[width=\columnwidth]{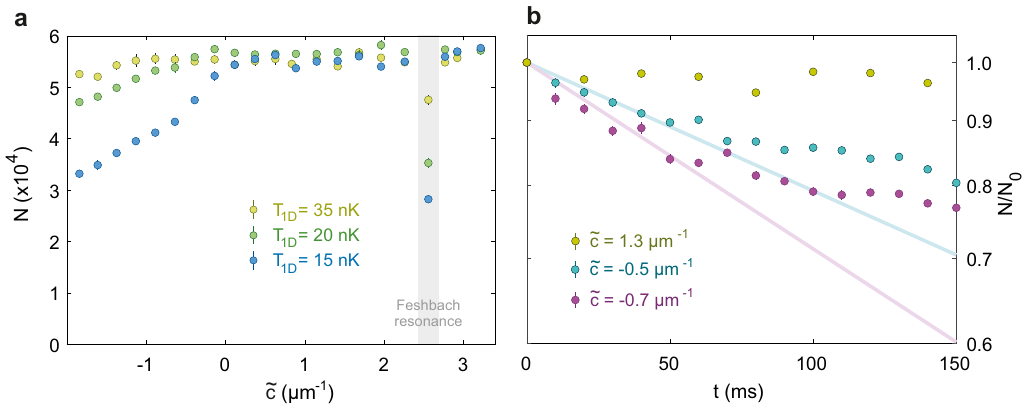}
	\caption{\textbf{Atom loss for different temperatures and interaction strengths.} \textbf{a}, Number of atoms after the interactions ramp as a function of $\cc$ for different temperatures. Blue, green, and yellow correspond to 1D temperatures $T_\text{1D}$ of $15$ nK, $20$ nK, $35$ nK, respectively. A Feshbach resonance is indicated by a gray shaded region. \textbf{b}, Evolution of normalized atom number for different values of $\cc$ at a temperature of $15$ nK. Each data point in is an average of five repetitions and the displayed error is equal to the standard error. The solid curves are exponential fits up to $50$ ms.  
	}
	\label{Fig_S5_Lifetime}
\end{figure*}
In our experiment, short-range molecular processes lead to atom loss. In Fig.~\ref{Fig_S5_Lifetime}\textbf{a} we show the atom losses during the interaction ramp corresponding to the measurements given in Fig.~\ref{Fig_3}\textbf{b}.
In the repulsive regime $\cc\!>\!0$, the atom number remains stable, however we see losses for interactions below $\cc\!=\!0$. We attribute this atom loss primarily to the formation of bound states, which enhance inelastic scattering processes not taken into account by the 1D Hamiltonian~\eqref{eq_ham}. Since we use Cs atoms in the lowest hyperfine ground state $|F,m_{F}\rangle=|3,3\rangle$, two-body inelastic processes are suppressed. Therefore, these losses in our experiment primarily originate from the three-body inelastic scattering.  The stability of the gas, depends both on the interaction strength and on the initial conditions of the protocol. The stronger the attraction, the tighter the bound states, with more frequent inelastic scattering. For a fixed $\cc$, the atom loss decreases at higher temperatures, as the probability of forming larger strings decreases at higher temperatures. While for $T_{\text{1D}}$=20 and 35 nK, the losses are less than 20\%, for the coldest $T_{\text{1D}}$=15 nK, we observe losses around 40\% for the strongest attractive interaction. Even with such high losses, the agreement between experimental data for the momentum distribution after 1D expansion (normalized with atom number) and GHD results as shown in Fig.~\ref{Fig_S5_Lifetime}\textbf{a} remains very good. Next, we measure the lifetime of the gas for different interaction strengths. In Fig.~\ref{Fig_S5_Lifetime}\textbf{b}, we show the normalized atom number as a function of hold time $t$ after the completion of the $\cc$ ramp for three different $\cc$ target values: $\cc\!=\!1.3, -0.5, -0.7$ $\mu$m$^{-1}$. For these measurements, we set $t_\text{1D}\!=\!0$ ms and $t_\text{3D}\!=\!36$ ms. In the repulsive regime, we do not observe any significant atom loss, however, in the attractive regime, we observe around 20\% atom loss over $t=150$ ms. Here, the density of the gas increases, resulting in higher three-body losses. Here we refrain from fitting the decay measurements with a three-body loss function and instead use a simple exponential function up to $50$ ms to estimate the lifetime. For $\cc = -0.5$ $\mu$m$^{-1}$, and $-0.7$ $\mu$m$^{-1}$, we obtain lifetime of $75.6(252)$ ms, and $57.4(178)$ ms respectively. These lifetimes are longer than the typical 1D expansion times $t_\text{1D}$ of 10 ms used in our experiment.

\section{Supplementary note 5: Simulating the Tonks-Girardeau regime}
\label{sec_supp3}

At strong repulsive interaction, the 1D Bose gas is well described by a gas of hard-core bosons, also known as Tonks-Girardeau (TG) regime and it is amenable of a straightforward theoretical treatment that we briefly recap~\cite{Girardeau1960}. Through a Jordan-Wigner transformation, hard-core bosons can be mapped into a system of free fermions. Hence, their hydrodynamics is that of free particles, satisfying the equation $\partial_t \rho_{t,z}(\lambda)+v(\lambda)\partial_z\rho_{t,z}(\lambda)-\partial_z V(z)\partial_k \rho_{t,z}(\lambda)=0$, with $v(\lambda)=\partial_\lambda E(\lambda)$ and $E(\lambda)=\tfrac{\hbar^2 \lambda^2}{2m}$ and $\rho_{t,z}(\lambda)$ being the local rapidity distribution. The initial condition to the hydrodynamic equations is given by thermal states, which we compute within the local density approximation  $\rho_{t=0,z}(\lambda)=\tfrac{1}{2\pi}\left(1+e^{\beta[ E(\lambda)-\mu+V(z)]}\right)^{-1}$, with the inverse temperature $\beta$ and the chemical potential $\mu$ being fitting parameters.

We compute the momentum distribution within a local density approximation (LDA). We fix the position $z$, and consider a homogeneous state described by the rapidity distribution $\rho_{t,z}(\lambda)$. On homogeneous states, one defines the fermionic correlator $F_z(\Delta z)=\int \dd\lambda\, e^{i\lambda \Delta z}\rho_{t,z}(\lambda)$, from which the bosonic correlation $g_{1,z}(\Delta z)\equiv \langle \hat{\psi}^\dagger(z+\Delta z/2)\hat{\psi}(z-\Delta z/2)\rangle $ can be computed, where we conveniently introduced the creation-annihilation bosonic fields $\hat{\psi}^\dagger(z)$ and $\hat{\psi}(z)$ obeying canonical commutation relations $[\hat{\psi}(z),\hat{\psi}^\dagger(z')]=\delta(z-z')$.
$g_{1,z}(\Delta z)$ is found within LDA by solving certain Fredholm integral equations~\cite{Lenard1966} (see also Ref.~\cite{Bastianello2017} for our notation) 
\be\label{eq_one_body}
g_{1,z}(\Delta z)=\text{det}[1-2 \mathcal{F}_{(\Delta z,0)}]\left[\mathcal{F}_{(\Delta z,0)}*(1-2\mathcal{F}_{(\Delta z,0)})^{-1}\right](\Delta z,0)\, ,
\ee
where $\mathcal{F}_{(\Delta z, 0)}$ is an operator acting on functions with support in $(\Delta z,0)$ and with entries $\mathcal{F}_{(\Delta z, 0)}(a,b)\equiv F_z(a-b)$. Its action on a test function $f$ is defines as $[\mathcal{F}_{b\in(\Delta z,0)}](a)\equiv \int_{(\Delta z,0)}\dd b\, \mathcal{F}_{(\Delta z, 0)}(a,b) f(b)$.
For brevity, we define the operator product on this space as $[A*B](a,b)\equiv \int_{y\in(\Delta z,0)}\dd y\, A(a,y)B(y,b)$. In Eq.~\eqref{eq_one_body}, after having computed the operator $\left[\mathcal{F}_{(\Delta z,0)}*(1-2\mathcal{F}_{(\Delta z,0)})^{-1}\right]$, one has to focus on its entries at the edge of the definition domain $\left[\mathcal{F}_{(\Delta z,0)}*(1-2\mathcal{F}_{(\Delta z,0)})^{-1}\right](\Delta z,0)$

In practice, determinants and matrix inversions are computed by discretizing the operator on a finite grid (about $120$ points in our simulations).
Finally, the momentum distribution is obtained by Fourier transform. More specifically, we define the LDA position-dependent momentum distribution as
\be
P_z(p)=\int \frac{\dd \Delta z}{2\pi}\, e^{-i p \Delta z}g_{1,z}(\Delta z)\, .
\ee
$P_z(p)$ describes the probability distribution in the momentum-space $(p,z)$ phase space in the longitudinal direction obtained right after the release of the three dimensional optical lattice: the final comparison with experimental data is obtained by further propagating the particles for a finite three dimensional TOF, assuming that $ P_z(p)$ describes the number of particles starting at $z$ and with velocity $v(p)=\tfrac{\hbar p}{m}$.

\section{Supplementary note 6: Characterizing the rapidity distribution in the TG regime}
\label{sec:supmat_Tonks_1D_expansion}

\begin{figure*}[t!]
\centering
\includegraphics[width=\textwidth]{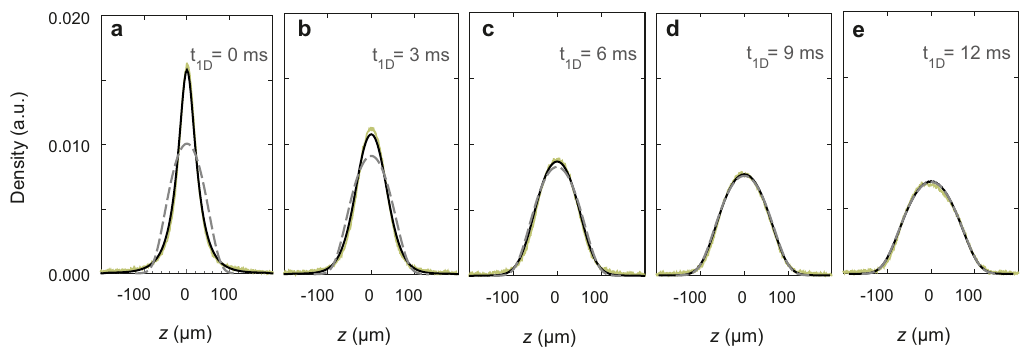}
	\caption{{\textbf{1D expansion in TG regime.}} \textbf{a} to \textbf{e}, TOF distributions of a TG gas following an expansion in 1D, with $t_\text{1D}\!=\!0 $ to $12$ ms. $z$ denotes the spatial position after TOF. The experimental data and simulation results are plotted in green, and black respectively. 
   The black curve is the density profile obtained expanding the true momentum distribution, whereas the grey dashed curve is obtained approximating the momentum distribution with the rapidity distribution. 
    Each experimental data is the average of five repetitions.
    }
\label{Fig:SupMat1_Tonks_expansion}
\end{figure*}

Theory predicts the longitudinal density profile of the expanded cloud after the system has entered a sufficiently dilute regime. Therefore, benchmarking the validity of this approximation is of key importance.
In the repulsive regime, the momentum distribution of a dilute gas is approximated by the rapidity distribution, but at finite density, interactions spoil this identification. Computing the momentum distribution at finite density and for arbitrary interaction strength is theoretically challenging, but it is accessible in the TG regime (see Supplementary Note~\ref{sec_supp3}). We use this to benchmark our experimental results.
We characterize our rapidity measurement protocol by first matching the experiment with the simulation in the TG regime for different 1D expansion times $t_\text{1D}$. In our experiment, after lattice loading, we ramp $a_\text{3D}$ from 210 $a_0$ to approximately 750 $a_0$ in $150$ ms. We then apply a $808$-nm anti-trapping beam that flattens out the longitudinal harmonic trap. We estimate that the longitudinal confinement of the tubes is flattened for a region of approximately $80$ $\mu$m around the center of the trap without affecting the transverse trapping. We allow the gas to expand in 1D for variable time $t_\text{1D}$, before taking a standard TOF absorption image. In Fig.~\ref{Fig:SupMat1_Tonks_expansion}, we show the TOF distribution of the gas after expansion in 1D for $t_\text{1D}\! =\! 0$ to $12$ ms. Throughout the 1D expansion, we see good agreement with the simulation results. For $t_\text{1D} = 0$ ms, the momentum and rapidity distributions are strikingly different.
As $t_\text{1D}$ increases, the distributions broaden and approach the rapidity distribution. After $t_\text{1D} = 6$ ms, the TOF distributions coincide with the rapidity distribution of the gas.

\section{Supplementary note 7: GHD and simulations at finite interactions}
\label{sec_supp4}

To give quantitative theoretical predictions for the finite interactions, we employ thermodynamic Bethe ansatz (TBA)~\cite{takahashi2005} to determine the initial state, and Generalized Hydrodynamics (GHD) to follow the time evolution.
Throughout this section, we use for simplicity adimensional units rescaling coordinates $z_i$ by a unit length $\ell=1$ $\mu\text{m}$, in such a way the Bose gas Hamiltonian reads 
\begin{equation}
\hat{H}=-\sum_i  \frac{\partial^2}{\partial z_i^2} +c\sum_{ i\ne j} \delta(z_i-z_j) + \sum_i V(z_i)\, ,
\end{equation}
with $c=\cc\,\ell$.

\bigskip

\textbf{Summary of notation.---}
It is convenient to define a unified notation. In integrable models, interactions are captured by ``dressing" bare quantities: for an arbitrary test function $\tau_n(\lambda)$, the dressing operation $\tau_n(\lambda)\to\tau_n^\dr(\lambda)$ is defined as the solution of the integral equation
\be\label{eq_dr}
\tau_n^\dr(\lambda)=\tau_{n}(\lambda)-\sum_{n'}\int \frac{\dd\lambda'}{2\pi}\varphi_{n,n'}(\lambda-\lambda')\vartheta_{n'}(\lambda')\tau_{n'}^\dr(\lambda')\, .
\ee
Above, the integral over the rapidities is on the whole real axis, and the summation over the internal index $n$ runs over the domain discussed below.
In Eq.~\eqref{eq_dr} one defines the filling fraction $\vartheta_n(\lambda)\equiv \rho_n(\lambda)/(2\pi (\partial_\lambda p_n(\lambda))^\dr$, with $p_n(\lambda)$ the bare momentum of the quasiparticle and $\varphi_{n,n'}(\lambda)$ the interaction-dependent scattering kernel. In the repulsive phase $c>0$ one has only terms for $n=1$, and $p_{n=1}(\lambda)=\lambda$, $\varphi_{n=1,n'=1}(\lambda)=-\frac{2c}{\lambda^2+c^2}$ and bare energy $E_{n=1}(\lambda)=\lambda^2$. In contrast, in the attractive case $n\in \mathbb{N}$ and one has $p_n(\lambda)=n\lambda$, $E_n(\lambda)=n\lambda^2-\tfrac{c^2}{12}n(n^2-1)$ and scattering kernel $\varphi_{n,n'}(\lambda)=(1-\delta_{n,n'})a_{|n-n'|}(\lambda)+2a_{|n-n'|+2}(\lambda)+2a_{|n-n'|+4}(\lambda)...+2a_{n+n'-2}(\lambda)+a_{n+n'}(\lambda)$, with $a_j(\lambda)=-\tfrac{4j c}{c^2 j^2+4\lambda^2}$. The thermodynamics and hydrodynamics of integrable models can be expressed in terms of these functions.
\bigskip

\textbf{Thermodynamics.---} The initial conditions of our experimental protocol within the repulsive phase are well-approximated by a thermal ensemble. Within the local density approximation, the filling function $\vartheta_{t=0,x}(\lambda)$ (where we suppress the $n-$label, since we focus on the repulsive phase $n=1$) is obtained by solving the following integral equations from thermodynamic Bethe ansatz~\cite{takahashi2005}
\be\label{eq_thermo}
\log[1/\vartheta_{t=0,z}(\lambda)-1]=\beta [E(\lambda)-\mu_\parallel+V(z)]-\int \frac{\dd\lambda'}{2\pi}\varphi(\lambda-\lambda')\log[1-\vartheta_{t=0,z}(\lambda')]\, .
\ee
When estimating the tube population in Supplementary Note~\ref{sec:Sup_mat_Exp_prep}, the chemical potential of each tube is renormalized by the transverse potential $\mu_\parallel\to\mu-V_\perp(x,y)$, where $V_\perp(x,y)$ is the transverse trapping potential in the $x-y$ plane. The global chemical potential $\mu$ and global inverse temperature $\beta$ are then considered fitting parameters.

\bigskip

\textbf{Hydrodynamics.---}We simulate the evolution with GHD, expressed in the space of filling fractions~\Cite{Alvaredo2016,Bertini2016} 
\be\label{eq_ghd_fill}
\partial_t \vartheta_n+v^\eff_n\partial_x\vartheta_n+F^\eff_n\partial_\lambda\vartheta_n=0
\ee
and equivalent to Eq.~\eqref{eq_ghd}, where we omitted all the variables for compactness. The GHD equations in the fillings' space are more stable for numerical purposes. The effective velocity can be computed as $v^\eff_n(\lambda)=(\partial_\lambda E_n)^\dr/(\partial_\lambda p_n)^\dr$, while the effective force $F^\eff_n(\lambda)=\mathcal{F}_n^\dr(\lambda)/(\partial_\lambda p_n)^\dr$ combines the effect of the trap, and of interaction changes in time $\mathcal{F}_n(\lambda)=-n \partial_z V+\partial_t c\sum_{n'}\int \tfrac{\dd\lambda'}{2\pi}\partial_c\Phi_{n,n'}(\lambda-\lambda')(\partial_{\lambda'}p_{n'})^\dr\vartheta_{n'}(\lambda')$, where one defines $\Phi_{n,n'}(\lambda)=\int^\lambda \dd\lambda'\, \varphi_{n,n'}(\lambda')$.
The above GHD equations describe the hydrodynamic evolution within the repulsive or attractive phase separately, but for passing from one to another the proper boundary conditions are needed. Physically, they describe how particles form Bethe strings by passing from the repulsive to the attractive phase. Using a maximum-entropy argument valid in the regime of slow interaction changes, these equations have been derived in Refs.~\cite{Koch2021,Koch_2022}
\be
\vartheta_{n;z}(\lambda)\Big|_{c\to 0^-}=1-\frac{\sinh(n\Omega_z(\lambda)/2)\sinh((n+2)\Omega_z(\lambda)/2)}{\sinh^2((n+1)\Omega_z(\lambda)/2)}\,
\ee
with $\Omega_z(\lambda)$ defined from the repulsive root density at vanishing interactions $\Omega_z(\lambda)=\lim_{c\to 0^+}\log\left[1+1/(2\pi \rho_{n=1;z}(\lambda))\right]$.
During the evolution, the density and energy profiles are computed in local density approximation.
An analytical expression for the local pair correlator has been obtained through the Hellmann-Feynmann theorem~\cite{Kheruntsyan2003,Bastianello2019}
\be\label{eq_psi2}
\langle [\hat{\psi}^\dagger(z)]^2[\hat{\psi}(z)]^2\rangle=\sum_n\int \dd\lambda\, \big\{\partial_c E_n(\lambda) \rho_n(\lambda)+\frac{1}{2\pi}\partial_\lambda E_n(\lambda)\vartheta_n(\lambda)f^\dr_n(\lambda)\big\}\, ,
\ee
whereas the total integrated pair correlation is $G_2\equiv \int \dd z\, \langle [\hat{\psi}^\dagger(z)]^2[\hat{\psi}(z)]^2\rangle$.
Above, $f_n(\lambda)=\sum_{n'}\int \dd\lambda'\, \Phi_{n,n'}(\lambda-\lambda')\rho_{n'}(\lambda)$, where, using a local density approximation, the TBA expressions are evaluated on the root density in position $z$.

\bigskip

\textbf{The momentum distribution after the one-dimensional asymptotic expansion.---} The analytical computation of the momentum-distribution of a dense Bose gas is a formidable challenge. However, in the approximation of a dilute gas obtained after a one-dimensional expansion from a dense gas, analytical results can be obtained.
In the repulsive case, isolated particles are well separated and the spatially resolved momentum distribution $P_{z}(p)$ coincides with the rapidity distribution $P_{z}(p)\simeq \rho_z(p)$~\cite{Caux2019}: to compare with the TOF experimental measurements, we assume that when the three dimensional optical trap is switched off, $P_z(p)$ is the density distribution of particles in position $z$ and with momentum $p$, which further freely evolve during the three-dimensional expansion. 

In the attractive regime, Bethe strings complicate the picture. We assume Bethe strings are spatially well-separated, and they independently contribute to the final momentum distribution. The momentum distribution of a single Bethe string is given by the modulus square of the overlap between the Bethe string's wavefunction and plane waves.
A Bethe string of rapidity $\lambda$ and $n$ particles has wavefunction $\psi_n(z_1<z_2<\dots<z_n)\propto e^{i  \sum_{j=1}^n(\lambda+ i\frac{c}{2}(n+1-2j)z_j}$, and the wavefunction is symmetrically extended to other orderings of the coordinates. Taking advantage of Galilean invariance and the scaling with the interaction $c$, we can write the total spatially-resolved momentum distribution as

\begin{figure*}[t!]
\centering
	\includegraphics[width=0.75\columnwidth]{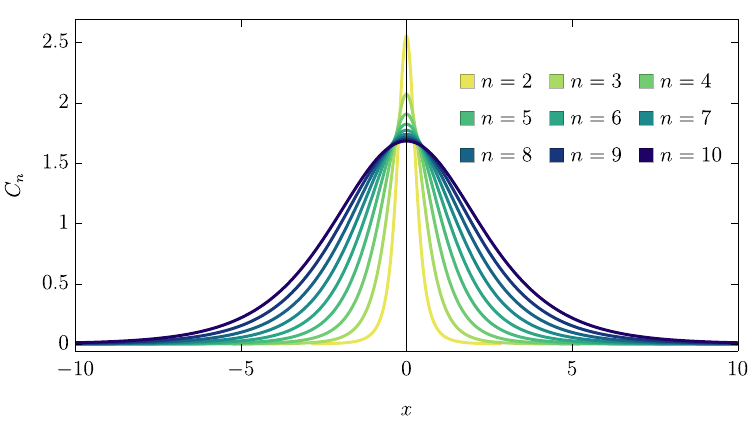}
	\caption{\textbf{The universal functions $C_n$.} We show the first universal functions appearing in the string-resolved momentum distribution after 1D expansion, see Eq.~\eqref{eq_Cn} and related discussion.
	}
	\label{Fig_SBS}
\end{figure*}

\be\label{eq_Cn}
P_z(p)=\sum_{n=1}^\infty \int \dd\lambda \frac{1}{|c|}C_n\left(\frac{p-\lambda}{|c|}\right)\rho_{n;z}(\lambda)\, ,
\ee
where the function $C_n$ captures the contribution of each string, and it is defined as the modulus squared of the overlap of the Bethe string wavefunction with zero real rapidity and computed at unit interaction $c=-1$, which we call $\bar{\psi}_n$
\be
C_n(x)=\int \dd z_1 e^{-i x z_1}\int \dd z_{j>1} \,\bar{\psi}_n^*(0,z_2,...,z_n)\bar{\psi}_n(z_1,z_2,...,z_n)\, .
\ee
Notice that above there is no restriction on the order of the coordinates. This integral can be simplified with some tedious combinatorics which can be carried over on a laptop and we overview below.
Modulus getting a factor $(n-1)!$, we can assume $z_2<z_3<...<z_n$ and the non-trivial permuations are now $z_1$ and $0$ with respect to the other coordinates: we expand the integral by summing over these domains. 
In the end, everything requires computing integrals in the form $I_j[q_1,...,q_j]=\int_{y_1<...<y_j<\infty} \dd^n y\, e^{i\sum_{a=1}^j q_j y_j}$ with imaginary $q_j$: by explicitly integrating the rightmost coordinate $y_j$, a simple recursive equation is found $I_j[q_1,...,q_{j-1}, q_j]=\frac{i}{q_j}I_{j-1}[q_1,...,q_{j-1}+q_j]$, which eventually gives a quick tabulation of $C_n(z)$. A Mathematica commented notebook that tabulates $C_n$ is provided on Zenodo~\cite{Zenodo}.
In Fig.~\ref{Fig_SBS} we show the first universal functions $C_n(z)$: they are bell-shaped functions normalized to the number of particles in the Bethe string $\int \dd z\, C_n(z)=n$ and of increasing width for larger $n$. 

\begin{figure*}[t!]
\centering
	\includegraphics[width=0.99\columnwidth]{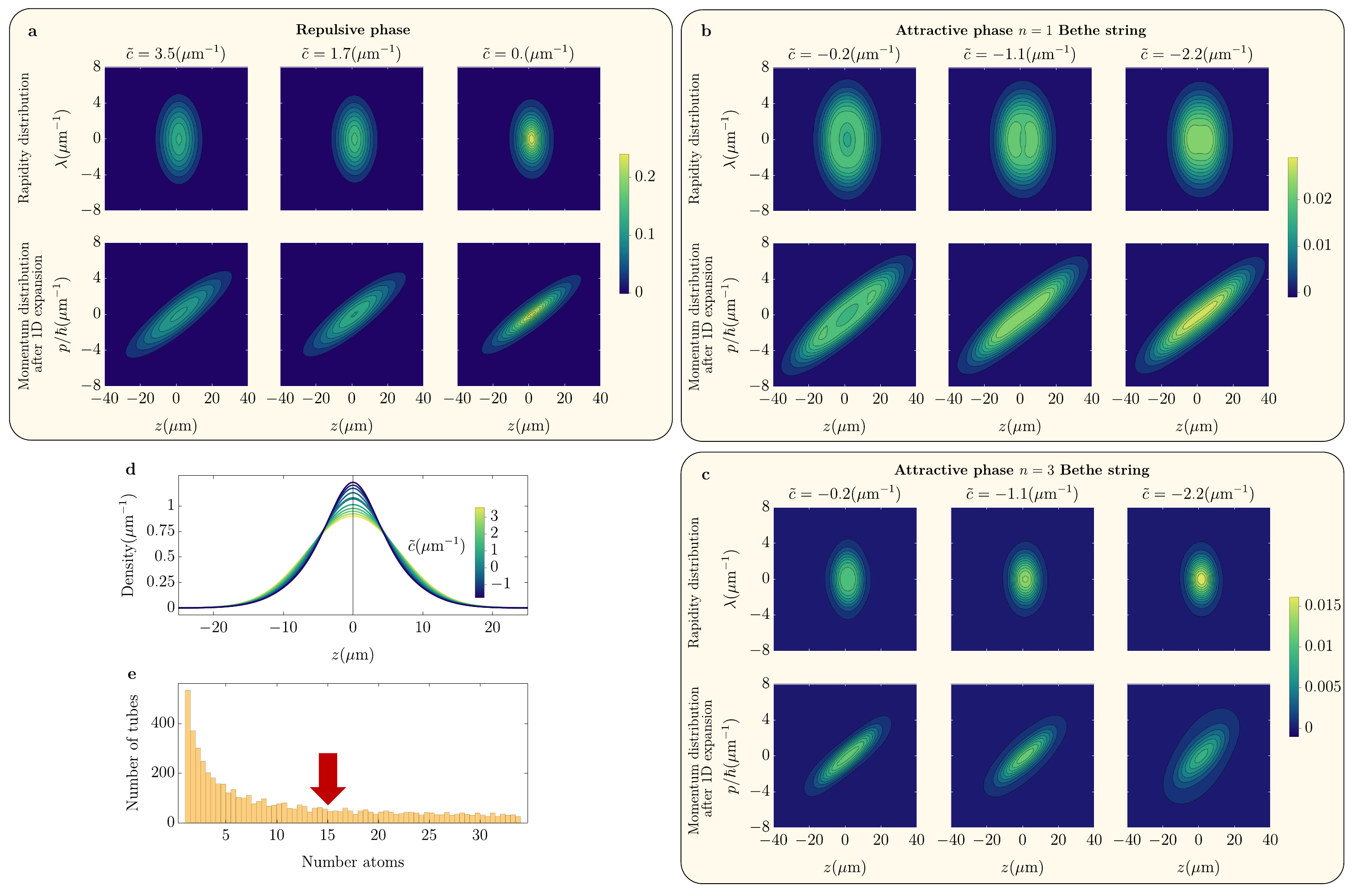}
	\caption{\textbf{Example of a typical evolution with GHD.} GHD evolution of a representative tube of 15 atoms in a harmonic trap with frequency 27.6 Hz, initially described by a thermal ensemble with temperature $T_{\text{1D}}=20$ nK. \textbf{a} to \textbf{c}, phase space density of the rapidity (first row) and momentum distribution per particle after 1D expansion of 10ms (second row). \textbf{a} shows the repulsive phase at different interactions, \textbf{b} to \textbf{c} focus on the example of the $n=1$ and $n=3$ Bethe strings. \textbf{d}, density profile in the harmonic trap, for different interactions. \textbf{e}, estimated atom number populations in the one-dimensional tubes for a crossover temperature of 10 nK.
	}
\label{Fig_SM_GHD}
\end{figure*}

\begin{figure*}[t!]
\centering
	\includegraphics[width=0.99\columnwidth]{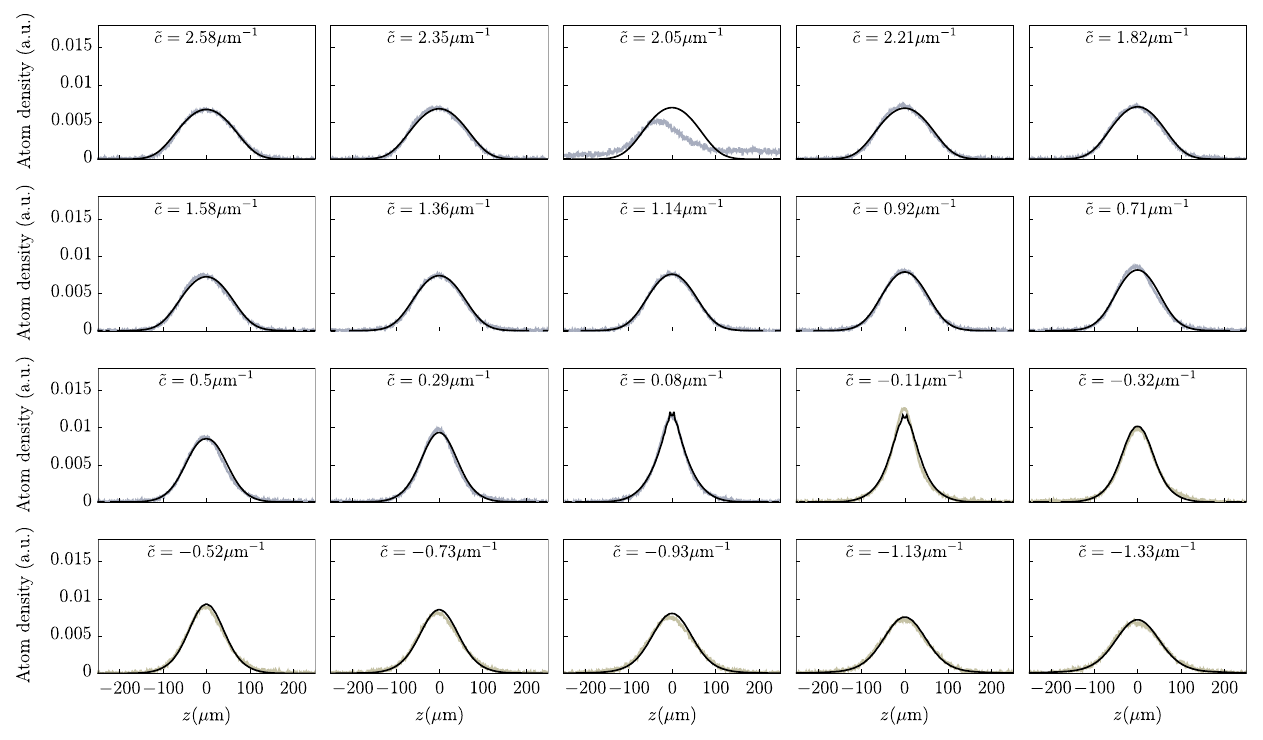}
\caption{\textbf{Comparison of experimental data with GHD.} For completeness, we show the full dataset of experimental data compared with the results of GHD simulations, for the dataset of Fig.~\ref{Fig_3} corresponding to $T_\text{1D}=15 \text{ nK}$. See caption of Fig.~\ref{Fig_3} for details. We show the atom density after a longitudinal expansion of $t_\text{1D}=10\text{ ms}$ and $t_\text{3D}=46.3 \text{ ms}$ in real space. 
Black solid lines are GHD curves, whereas colored lines are experimental data (repulsive: light blue, attractive: light green). 
The starting point of the interaction ramp is at $\cc=2.58 \,\mu\text{m}^{-1}$, which is then linearly changed in time.
The discrepancy shown at $\cc=2.05 \,\mu\text{m}^{-1}$ is attributed to the fact that it is close to a Feshbach resonance. When ramping the magnetic field to reach less repulsive interactions and the attractive phase, we jump across the resonance to obtain a smooth interaction ramp. The small dents shown by GHD curves at the weakest repulsive and attractive interactions are numerical artifacts arising during the extraction of the 1D-expanded profile from the filling function. This can be reduced by improving the discretization.
	}
\label{Fig_SM_many}
\end{figure*}
While the full profile for $C_n(z)$ is tedious to recover analytically for large $n$, the variance is rather easy to compute.
We consider directly the momentum variance $\langle p^2\rangle_{n,\lambda}$ of a Bethe string of rapidity $\lambda$ and $n$ components. We can leverage on the knowledge of the energy of this state $E_n(\lambda)=\langle H\rangle_{n,\lambda}=n\lambda^2-\frac{c^2}{12}n(n^2-1)$. From the Hellmann-Feynman theorem we know $\partial_c E_n(\lambda)=\langle \partial_c H\rangle_{n,\lambda}$, and with the explicit observation that $\partial_c H=\, \sum_{i\ne j}\delta(z_i-z_j)$ we observe 
\be
\langle p^2\rangle_{n,\lambda}=\langle H-c\sum_{i,j}\delta(x_i-x_j)\rangle_{n,\lambda}=E_n(\lambda)-c\partial_c E_n(\lambda)= n\lambda^2+\frac{c^2}{12} n(n^2-1)\, .
\ee
Specifying $\lambda=0$ and $c=-1$, we connect with the functions $C_n$ getting $\int \dd y \,C_n(y)=\tfrac{1}{12}n(n^2-1)$.
The total variance of the momentum distribution after 1D expansion is obtained by summing over all the strings with the proper weight $\rho_n(\lambda)$. 
\bigskip

\textbf{Numerical discretization.---} The filling functions appearing in the GHD equations~\eqref{eq_ghd_fill} are discretized on a finite grid in the rapidity and real space. The main bottleneck for the simulations is the solution of the (discretized) integral equations defining the dressing operation, whose matrices grow with the number of strings times the number of points in the rapidity discretization. We use $100$ points in the space discretization, $50$ points in the rapidity's grid for a maximum number of $25$ strings for Fig.~\ref{Fig_3}, and increase it to $30$ strings for Fig.~\ref{Fig_4} where we did not numerically expanded in 1D, but directly computed the energies from the GHD. For the time-evolution of GHD, we use the method of characteristics with the second order implementation described in Ref.~\cite{Bastianello2019}.
In Fig.~\ref{Fig_SM_GHD}, we provide further details on a typical simulation focusing on theory only. For the sake of concreteness, we focus on one of the datasets discussed in Fig.~\ref{Fig_3}, more precisely we consider the case $T_\text{1D}=20$ nK, and we focus on a single representative tube with 15 atoms. In Fig.~\ref{Fig_SM_GHD}\textbf{a} to~\textbf{c}, we show density plots of the rapidity distribution in the harmonic trap, and the spatially-resolved momentum distribution after 1D expansion of 10 ms, for different interaction strengths. Figure~\ref{Fig_SM_GHD}\textbf{a} shows the repulsive side (no strings), whereas in Fig.~\ref{Fig_SM_GHD}\textbf{b} and~\textbf{c}, we consider the attractive phase and focus, as an example, on the $n=1$ Bethe string (one particle) and $n=3$ Bethe string (three-particle bound state) respectively. As the interaction $\cc$ is changed from strongly to weakly repulsive, and then from weakly attractive to strongly attractive the rapidity distribution gets squeezed in the center of the trap. Upon 1D expansion, the momentum distribution in the repulsive branch and of the $n=1$ Bethe string are simply a deformation of the rapidity distribution obtained ballistically propagating the particles in the $z$-direction, the magnitude of the velocity increases with the momentum giving the apparent rotation of the oval. For the $n=3$ Bethe string, this effect is superimposed with the energy release upon breaking the bound states, that dilates the distribution in the vertical direction. This effect is stronger at larger values of attractive interaction. In Fig.~\ref{Fig_SM_GHD}\textbf{d}, we show the density profile within the harmonic trap, where the squeezing in the center of the trap is evident. Finally, in Fig.~\ref{Fig_SM_GHD}\textbf{e} we provide the estimated atom population across the tubes (see Supplementary Note~\ref{sec:Sup_mat_Exp_prep}). The dimensional-crossover temperature that best matches experimental data is 10 nK.
For completeness, in Fig.~\ref{Fig_SM_many} we show a full comparison of the experimental expanded cloud with results of GHD simulations for the dataset corresponding to $T_{1D}=15 \text{ nK}$ shown in Fig.~\ref{Fig_3}.

\bigskip

\section{Supplementary note 8: The virial theorem and Generalized Hydrodynamics}
\label{sec_S5}

\noindent The virial theorem 
\be
2U-2E_V-E_\text{int}=0
\ee
is well-known for equilibrium states~\cite{tan2008generalized}. In this section, we prove its validity for non-equilibrium states that are stationary with respect to the GHD equations.
We work in the adimensional units of Supplementary Note~\ref{sec_supp4} and start by manipulating the TBA expression for the pair correlator~\eqref{eq_psi2}.
 We consider the attractive phase, but the same calculations hold in the repulsive case considering  the $n=1$ case only. In the Lieb-Liniger model, the scattering kernel is not an independent function of $\lambda$ and $c$, but a scaling function of their ratio $\partial_c \Phi_{n,n'}(\lambda)=-\tfrac{\lambda}{c}\partial_\lambda \Phi_{n,n'}(\lambda)=-\tfrac{\lambda}{c} \varphi_{n,n'}(\lambda)$. Using this fact, and the symmetry of the kernel and of the dressing function $\sum_n\int \dd\lambda \vartheta_n(\lambda)a_n(\lambda)b^\dr_n(\lambda)=\sum_n\int \dd\lambda \vartheta_n(\lambda)a^\dr_n(\lambda)b_n(\lambda)$ for any function $a_n(\lambda)$ and $b_n(\lambda)$, with straightforward manipulations we can write
\be
\langle [\hat{\psi}^\dagger(z)]^2[\hat{\psi}(z)]^2\rangle=\frac{1}{c}\sum_n \int \dd\lambda\, \left[\partial_c E_n(\lambda)+\lambda\partial_\lambda E_n(\lambda)\right]\rho_n(\lambda)+\frac{1}{c}\sum_n \int \dd\lambda\, \lambda \partial_\lambda p_n(\lambda)(\partial_\lambda E_n)^\dr\vartheta_n(\lambda)\, .
\ee
\begin{figure*}[t!]
\centering
	\includegraphics[width=0.5\columnwidth]{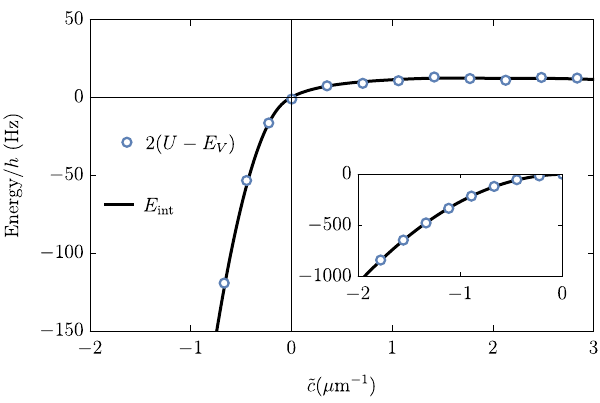}
	\caption{\textbf{Benchmark of the virial theorem in GHD simulations.} For the same dataset provided in Fig.~\ref{Fig_4}, but focusing on theory only, we check that the corrections to the virial theorem due to the non-stationary nature of the state and to the interaction changes, as discussed in Eq.~\eqref{eq_S14}, are negligible. The black line shows the total interaction energy obtained in the local density approximation from Eq.~\eqref{eq_S12}. Markers are twice the difference between the internal and potential energies, and agree well with the interaction energy: the small oscillations shown in the repulsive phase are due to the corrections discussed in Eq.~\eqref{eq_S14}, and are negligible compared to the experimental uncertainty. In the inset, we focus on the strongly attractive regime, the axis of the inset are same of the main plot, but on a different scale: in this regime, the binding energy of Bethe strings dominates the virial theorem. In these simulations, we considered $20$ strings, in contrast with Fig.~\ref{Fig_4} where $30$ strings have been used. Since the virial theorem holds for any filling fraction stationary to the GHD equations, the truncation in the number of strings does not spoil the agreement between the two curves.
	}
	\label{Fig_SM_virial}
\end{figure*}

We now use the explicit expressions for the energy and momentum in the Lieb-Liniger model, and notice $\partial_c E_n(\lambda)+\lambda\partial_\lambda E_n(\lambda)=2E_n(\lambda)$  and $\lambda\partial_\lambda p_n(\lambda)=p_n(\lambda)$. Multiplying both sides by $c$ and integrating over $z$, we recognize the total interaction energy $E_\text{int}=c\int \dd z\, \langle [\hat{\psi}^\dagger(z)]^2[\hat{\psi}(z)]^2\rangle $ and total internal energy $U=\int \dd z\, \sum_n\int \dd\lambda\, E_n(\lambda) \rho_n(\lambda)$, where the $z-$dependence of the root density and filling fraction is left implicit for the sake of notation.
\be\label{eq_S12}
 E_\text{int}=2 U+\int \dd z\,\sum_n \int \dd\lambda\,  p_n(\lambda)(\partial_\lambda E_n)^\dr\vartheta_n(\lambda)\, .
\ee
So far we have not used the GHD equations yet, but only the local density approximation and symmetries of the TBA. The last step uses the GHD equations to further manipulate the remaining integral in Eq.~\eqref{eq_S12}. In particular, it is possible to show
\begin{multline}\label{eq_S14}
\int \dd z\,\sum_n \int \dd\lambda\,  p_n(\lambda)(\partial_\lambda E_n)^\dr\vartheta_n(\lambda)=- \langle z \partial_z V\rangle-\frac{\dd}{\dd t}\left(\sum_n \int \dd z \int \dd \lambda\,  z p_n(\lambda) \rho_n(\lambda)\right)+\\
+\partial_t c\int \dd z\sum_n\int \frac{\dd\lambda}{2\pi} z n\left(\sum_{n'}\int \dd\lambda'\partial_c\Phi_{n,n'}(\lambda-\lambda')\rho_{n'}(\lambda')\right)\, .
\end{multline}
The proof is technical and reported below. If the potential $V(z)$ is harmonic, then $z\partial_z V(z)=2V(z)$ and one recovers the potential energy $\langle z\partial_zV\rangle=2 E_V$. In Eq. \eqref{eq_S14}, the total derivative is reminiscent of the virial theorem for classical particles, and there is a further contribution coming from interaction changes. If we consider stationary states, then $\partial_t c=0$ and the total derivative vanishes as well, thus we recover the virial theorem $E_\text{int}-2U+2E_V=0$. The virial theorem remains a good approximation for slowly evolving states. In our simulations, we explicitly checked that the interaction energy computed through the virial theorem agrees with the internal energy from \eqref{eq_psi2}, see Fig.~\ref{Fig_SM_virial}, and both agree well with the experimental data as shown in Fig.~\ref{Fig_4}.\\

\textbf{Proof of Eq.~\eqref{eq_S14}.---}
To prove Eq.~\eqref{eq_S14}, we first go through a convenient change of variables in the GHD equations. More precisely, we move from the $(z,\lambda)$ plane to a pair of conjugated canonical variables $(z,\Dp)$. We define
\begin{eqnarray}\label{eq_Dp}
\Dp_n(\lambda)&=&p_n(\lambda)-\sum_{n'}\int \frac{\dd\lambda'}{2\pi}\Phi_{n,n'}(\lambda-\lambda')\vartheta_{n'}(\lambda')(\partial_{\lambda'}p_{n'}(\lambda'))^\dr\\
\Den_n(\lambda)&=&E_n(\lambda)+n V(z)-\sum_{n'}\int \frac{\dd\lambda'}{2\pi}\Phi_{n,n'}(\lambda-\lambda')\vartheta_{n'}(\lambda')(\partial_{\lambda'}E_{n'}(\lambda'))^\dr\label{eq_Den}
\end{eqnarray}
And consider the string-dependent change of variable $\lambda\to \Dp=\Dp_n(\lambda)$, likewise we define the filling fraction in the new space $\bar{\vartheta}_{n,z}(\Dp)$ through the identity $\bar{\vartheta}_{n,z}(\Dp_n(\lambda))=\vartheta_n(\lambda)$. We furthermore define $\mathcal{H}_n(z,\Dp)$ as $\mathcal{H}_n(z,\Dp_n(
\lambda)
)=\Den_n(\lambda)$, where the $z-$dependence is implicit in the integral~\eqref{eq_Den}.

We now rewrite the GHD equations in the filling fraction space~\eqref{eq_ghd_fill} in terms of the new variables.
Notice that $\partial_\lambda\Dp_n(\lambda)=(\partial_\lambda p_n)^\dr$ and $\partial_\lambda\Den_n(\lambda)=(\partial_\lambda E_n)^\dr$, therefore $\tfrac{1}{(\partial_\lambda p_j)^\dr}\partial_\lambda\to \partial_\Dp$ and $v^\text{eff}_n(\lambda)\to \partial_\Dp\mathcal{H}_n(x,\Dp)$. Using these identities in Eq.~\eqref{eq_ghd_fill} one obtains $(\partial_t+\partial_t\Dp_j\partial_{\Dp})\bar{\vartheta}_n(\Dp)+\partial_{\Dp}\mathcal{H}_n\partial_x\bar{\vartheta}_n(\Dp)+\mathcal{F}^\dr\partial_{\Dp}\bar{\vartheta}_n(\Dp)=0$, where it must be stressed that the change of variable $\lambda\to \Dp$ is time dependent since Eq.~\eqref{eq_Dp} depends on the evolving state. As a next step, one shows $\partial_t\Dp_n+\partial_z \Den_n=-\mathcal{F}^\dr_n$ through straightforward manipulations. First, one takes the time and space derivatives of both sides of Eqs.~\eqref{eq_Dp} and~\eqref{eq_Den} respectively, and sums the two equations.
The GHD equations in the space of root densities $\partial_t[(\partial_\lambda p_n)^\dr \vartheta_n(\lambda)]+\partial_x( (\partial_\lambda E_n)^\dr \vartheta_n(\lambda)]+\partial_\lambda[ 
\mathcal{F}^\dr\vartheta_n(\lambda)]=0$ (notice that $\rho_n(\lambda)=\tfrac{1}{2\pi}(\partial_\lambda p_n)^\dr \vartheta_n(\lambda)$) are used for further simplifications. With this last step, one obtains $\partial_t\Dp_n+\partial_z \Den_n=-\mathcal{F}_n(\lambda)+\sum_{n'}\int \dd\lambda'\, \Phi_{n,n'}(\lambda-\lambda')\vartheta_{n'}(\lambda')[-\mathcal{F}^\dr_{n'}(\lambda')]$, and thus can identify $\partial_t\Dp_n+\partial_z \Den_n=-\mathcal{F}^\dr_n$ as anticipated. Using now $\partial_z\Den_n\to \partial_z \mathcal{H}_{n}(x,\Dp)$, one can rewrite the GHD equations in an explicit symplectic form
\be
\partial_t\bar{\vartheta}_{n,x}(\Dp)+\partial_{\Dp}\mathcal{H}_j(x,\Dp)\partial_x\bar{\vartheta}_{n,x}(\Dp)-\partial_x\mathcal{H}_n(x,\Dp)\partial_{\Dp}\bar{\vartheta}_{n,x}(\Dp)=0\, .
\ee
These are the Liouville's equations~\cite{arnol2013} in the phase space $(z,\Dp)$ for a particle evolving with classical Hamiltonian $\mathcal{H}_n(z,\Dp)$. The evolution can be seen in two ways: either as a function evolving in a time-independent phase space, or as if the coordinates are evolving with a fixed density background $\bar{\vartheta}_{n;t,z(t)}(\Dp(t))=\bar{\vartheta}_{n;t=0,z}(\Dp)$, where coordinates obey the equations of motion $\dot{z}=\partial_\Dp \mathcal{H}_n$ and $\dot{\Dp}=-\partial_z \mathcal{H}_n$. To prove Eq.~\eqref{eq_S14}, the second interpretation is more convenient. Notice that the Liouville's theorem~\cite{arnol2013} guarantees that the phase-space volume $\dd z\dd \Dp$ is constant in time. We consider the left-hand side of Eq.~\eqref{eq_S14} and express it in the $(z,\Dp)$ plane $\sum_{n}\int\dd z\dd\Dp \, \frac{ p_n(\lambda_n(\Dp))}{2\pi}\partial_\Dp \mathcal{H}_n\bar{\vartheta}_n(\Dp)$, where $\lambda_n(\Dp)$ denotes the inverse change of variable $\lambda\to \Dp$. We use $\partial_{\Dp}\mathcal{H}_n=\dot{z}$ and straightforward identities reaching
\be\label{eq_S19}
\int \dd z\,\sum_n \int \dd\lambda\,  p_n(\lambda)(\partial_\lambda E_n)^\dr\vartheta_n(\lambda)=\frac{\dd }{\dd t}\left(\sum_{n}\int\dd z\dd\Dp \, \frac{ p_n(\lambda_n(\Dp))}{2\pi}z\bar{\vartheta}_n(\Dp)\right)-\sum_{n}\int\frac{\dd z\dd\Dp}{2\pi} \, \frac{\dd  p_n(\lambda_n(\Dp))}{\dd t}z\bar{\vartheta}_n(\Dp)\, .
\ee
We now use the chain rule to compute $\frac{\dd  p_n(\lambda_n(\Dp))}{\dd t}=\partial_t [p_n( \lambda_n(\Dp))]\Big|_{\text{fixed }\Dp}+\dot{\Dp}\partial_\Dp  (p_n \lambda_n(\Dp))$, then we use $\dot{\Dp}=-\partial_z \mathcal{H}_n$ and move to time derivatives at fixed $\lambda$ by using $\partial_t [p_n( \lambda_n(\Dp))]\Big|_{\text{fixed }\Dp}=\partial_t p_n(\lambda_n(\Dp))-\partial_t \Dp_n\partial_\Dp [p_n( \lambda_n(\Dp))]$. Therefore, we can write $\frac{\dd  p_n(\lambda_n(\Dp))}{\dd t}=\partial_t p_n(\lambda_n(\Dp))+\mathcal{F}_n^\eff(\lambda_n(\Dp))\partial_\Dp  (p_n( \lambda_n(\Dp)))$. We notice that, in the Lieb Liniger model, $\partial_t p_n(\lambda)=0$. Using this last identity in Eq.~\eqref{eq_S19} and rewriting it in terms of the rapidities, Eq.~\eqref{eq_S14} follows.

\end{document}